\documentclass[english,superscriptaddress,nofootinbib]{revtex4}
\usepackage{amsmath}
\usepackage{makeidx}
\usepackage{amssymb}
\usepackage[T1]{fontenc}
\usepackage[latin9]{inputenc}
\usepackage{graphicx}
\usepackage{babel}

\makeatletter
\@ifundefined{textcolor}{}
{
 \definecolor{BLACK}{gray}{0}
 \definecolor{WHITE}{gray}{1}
 \definecolor{RED}{rgb}{1,0,0}
 \definecolor{GREEN}{rgb}{0,1,0}
 \definecolor{BLUE}{rgb}{0,0,1}
 \definecolor{CYAN}{cmyk}{1,0,0,0}
 \definecolor{MAGENTA}{cmyk}{0,1,0,0}
 \definecolor{YELLOW}{cmyk}{0,0,1,0}
 }
\makeatother

\begin{document}

\title{Shear and Bulk Viscosities of a Gluon Plasma in Perturbative QCD:
Comparison of Different Treatments for the $gg\leftrightarrow ggg$ Process}
\author{Jiunn-Wei Chen}
\affiliation{Department of Physics, Center for Theoretical Sciences, and Leung Center for
Cosmology and Particle Astrophysics, National Taiwan University, Taipei
10617, Taiwan}
\author{Jian Deng}
\affiliation{School of Physics, Shandong University, Shandong 250100, People's Republic
of China}
\author{Hui Dong}
\affiliation{School of Physics, Shandong University, Shandong 250100, People's Republic
of China}
\author{Qun Wang}
\affiliation{Interdisciplinary Center for Theoretical Study and Department of Modern
Physics, University of Science and Technology of China, Anhui 230026,
People's Republic of China}

\begin{abstract}
The leading order contribution to the shear and bulk viscosities, $\eta $ 
and $\zeta $, of a gluon plasma in perturbative QCD includes the 
$gg\leftrightarrow gg$\ (22) process, $gg\leftrightarrow ggg$\ (23) process
and multiple scattering processes known as the Landau-Pomeranchuk-Migdal
(LPM) effect. Complete leading order computations for $\eta $ and $\zeta $ 
were obtained by Arnold, Moore and Yaffe (AMY) and Arnold, Dogan and Moore
(ADM), respectively, with the inelastic processes computed by an effective 
$g\leftrightarrow gg$ gluon splitting. We study how complementary
calculations with 22 and 23 processes and a simple treatment to model the
LPM effect compare with the results of AMY and ADM. We find that our results agree
with theirs within errors. By studying the contribution of the 23 process to $\eta $, 
we find that the minimum angle $\theta$ among the final state gluons  
in the fluid local rest frame 
has a distribution that is peaked at $\theta \sim \sqrt{\alpha _{s}}$, 
analogous to the near collinear splitting asserted by AMY and ADM. However,
the average of $\theta $\ is much bigger than its peak value, as its
distribution is skewed with a long tail. The same $\theta $  behavior is also 
seen if the 23 matrix element is taken to the soft gluon bremsstrahlung
limit in the center-of-mass (CM) frame. This suggests that the soft
gluon bremsstrahlung in the CM frame still has some near collinear behavior in the
fluid local rest frame. We also generalize our result to a general $SU(N_{c})
$ pure gauge theory and summarize the current viscosity computations in QCD.
\end{abstract}

\maketitle

\section{Introduction}

Shear and bulk viscosities, $\eta $ and $\zeta $, are transport coefficients
characterizing how fast a system goes back to equilibrium under a shear mode
perturbation and a uniform expansion, respectively. In a weakly interacting
hot gluon plasma, $\eta $ is inversely proportional to the scattering rate, $%
\eta \propto 1/\Gamma \propto 1/\alpha _{s}^{2}\ln \alpha _{s}^{-1}$ \cite%
{Arnold:2000dr}, where $\alpha _{s}$ is the strong coupling constant. $\zeta 
$ is suppressed by an additional factor of $\left( T_{\mu }^{\mu }\right)
^{2}$, arising from the response of the trace of the energy momentum tensor $%
\left( T_{\mu }^{\mu }\right) $\ to a uniform expansion. Thus, $\zeta $
vanishes when the system is \textquotedblleft conformal\textquotedblright\
or scale invariant. For a gluon plasma, the running of the coupling constant
breaks the scale invariance. Thus, $T_{\mu }^{\mu }\propto \beta \left(
\alpha _{s}\right) \propto \alpha _{s}^{2}$, $\zeta \propto \alpha
_{s}^{2}/\ln \alpha _{s}^{-1}$ \cite{Arnold:2006fz}. In the perturbative
region, $\zeta /\eta \propto \alpha _{s}^{4}\ll 1$.

In the strong coupling region, smaller $\eta $\ is expected. The so-called
\textquotedblleft perfect fluid\textquotedblright\ is a fluid with the
smallest shear viscosity per entropy density ($s$) ratio, $\eta /s$. It is
conjectured that $\eta /s$ has a minimum bound $1/(4\pi )$ \cite%
{Kovtun:2004de}. This is motivated by the uncertainty principle of quantum
mechanics because $\eta /s$ is related to $\Delta E\Delta t$, the mean
energy and life time of quasiparticles. While the number $1/(4\pi )$ arises
from the universal value $\eta /s=1/(4\pi )$ obtained for a big class of
strongly interacting conformal field theories (CFT's) in the large $N$, $N$
being the size of the gauge group, and large t'Hooft coupling limits \cite%
{Kovtun:2004de,Buchel:2003tz,Buchel:2004qq}. This class of strongly
interacting CFT's are dual to another class of weakly interacting
gravitational theories in anti-de-Sitter space backgrounds. This
anti-de-Sitter space/conformal field theory correspondence (AdS/CFT) \cite%
{Maldacena:1997re,Gubser:1998bc,Witten:1998qj} allows that $\eta /s$\ in
strongly interacting CFT's can be computed in weakly interacting
gravitational theories.

The smallest $\eta /s$ known so far is realized in the hot and dense matter
(thought to be a quark gluon plasma of QCD) just above the phase transition
temperature ($T_{c}$) produced at RHIC \cite%
{Arsene:2004fa,Back:2004je,Adcox:2004mh} with $\eta /s=0.1\pm 0.1(\mathrm{%
theory})\pm 0.08(\mathrm{experiment})$ \cite{Luzum:2008cw}. A robust upper
limit $\eta /s<5\times 1/(4\pi )$ was extracted by another group \cite%
{Song:2008hj} and a lattice computation of gluon plasma yields $\eta
/s=0.102(56)$ (at temperature $T=1.24T_{c}$) \cite{Meyer:2007ic}. Away from $%
T_{c}$, $\eta /s$ of QCD becomes larger due to small couplings at high $T$\
or small derivative Goldstone boson couplings at low $T$. We will summarize
the current status of QCD $\eta /s$ vs. $T$ in Fig. \ref{fig:Tc}.

As for the bulk viscosity $\zeta $, it is small in the perturbative region.
However, near $T_{c}$, the rapid change of degrees of freedom gives a
rapid change of $T_{\mu }^{\mu }$ which could give very large $\zeta /s$ 
\cite{Kharzeev:2007wb,Karsch:2007jc}.

The best perturbative QCD calculation of $\zeta $ was carried out by Arnold,
Dogan and Moore (ADM) \cite{Arnold:2006fz} using the same approach as the 
$\eta $ computed by Arnold, Moore and Yaffe (AMY) in Refs. 
\cite{Arnold:2000dr,Arnold:2003zc}. In both $\eta $ and $\zeta $, the leading
order (LO) contribution involves the elastic process $gg\leftrightarrow gg$ (22),
inelastic number changing process $gg\leftrightarrow ggg$ (23) and
multiple scattering processes known as the Landau-Pomeranchuk-Migdal (LPM)
effect. In the complete leading order computations for $\eta $ and $\zeta $
obtained by AMY and ADM, respectively, the inelastic processes were
computed using an effective $g\leftrightarrow gg$ gluon splitting obtained
after solving sophisticated integral equations. 

In this paper, we study how complementary calculations with the 22 and 23
processes and a simple treatment to model the LPM effect compare with the
results of AMY and ADM. This approach is similar to the one used by Xu and
Greiner (XG) \cite{Xu:2007ns,Wesp:2011yy} who claimed that the dominant
contribution to $\eta $\ is 23 instead of 22, in sharp contradiction to the
result of AMY. 
While our approach is not model independent due to our simplified treatment
of the LPM effect, it can be used to double check XG's result since the two
approaches are very similar. We find that we cannot reproduce XG's result
unless the 23 collision rate is at least multiplied by a factor 6 (part of
this result was asserted in Ref.\ \cite{Chen:2010xk}). In the mean time, 
our $\eta $ agrees with AMY's within errors, while our $\zeta $ also agrees
with ADM's within errors.

Although our result does not provide a model independent check to AMY and
ADM's results, we can still study the angular correlation between final
state gluons using our approach. Because the 23 matrix element that we use
is exact in vacuum, we can check, modulo some model dependent medium effect,
whether the correlation is dominated by the near collinear splitting as
asserted by AMY and ADM. 

We study the distribution of the minimum angle $\theta $ among the final
state gluons. If the near collinear splittings dominate,
then most probable configurations would be that two gluons' directions  
are strongly correlated and their relative angle tends to be 
the smallest among the three relative angles in
the final state. This can be seen most easily in the center-of-mass (CM)
frame of the 23 collision with two gluons going along about the same
direction while the third one is moving in the opposite direction. We
expect it is also the case in the fluid local rest frame. 

We find that the distribution of $\theta $ is peaked at $\theta \sim \sqrt{\alpha _{s}}$, 
analogous to the near collinear splitting asserted by AMY and ADM. However,
the average of $\theta $, $\left\langle \theta \right\rangle $, is much
bigger than its peak value, as its distribution is skewed with a long tail.

The same $\theta $ behavior is also seen if the 23 matrix element is taken to the
soft gluon bremsstrahlung limit in the CM frame. This suggests that 
the soft gluon bremsstrahlung in the CM frame still has some near collinear 
behavior in the fluid local rest frame. 

We also generalize our result to a general $SU(N_{c})$\ pure gauge theory
and summarize the current viscosity computations in QCD.

\section{Kinetic theory with the 22 and 23 processes}

In this section, we will focus on the $\zeta $ computation. We refer the
formulation for calculating  $\eta $ to Ref.\ \cite{Chen:2010xk}. 

Using the Kubo formula, $\zeta $ can be calculated through the linearized
response function of a thermal equilibrium state $\left\vert \Omega
\right\rangle $ 
\begin{equation}
\zeta =\lim_{\omega \rightarrow 0}\frac{1}{9\omega }\int_{0}^{\infty }dt\int
d^{3}x \,e^{i\omega t}\,\langle \Omega \left\vert \lbrack T_{\mu }^{\mu
}(x),T_{\nu }^{\nu }(0)]\right\vert \Omega \rangle \,.
\end{equation}%
In the LO expansion of the coupling constant, the computation involves an
infinite number of diagrams \cite{Jeon:1994if,Jeon:1995zm}. However, it is
proven that the summation of the LO diagrams in a weakly coupled $\phi ^{4}$
theory \cite%
{Jeon:1994if,Jeon:1995zm,Carrington:1999bw,Wang:1999gv,Hidaka:2010gh} or in
hot QED \cite{Gagnon:2007qt} is equivalent to solving the linearized
Boltzmann equation with temperature-dependent particle masses and scattering
amplitudes. This conclusion is expected to hold in perturbative QCD as well.

The Boltzmann equation of a hot gluon plasma describes the evolution of the
color and spin averaged gluon distribution function $f_{p}(x)$ which is a 
function of space-time $x=(t,\mathbf{x})$ and momentum $p=(E_{p},\mathbf{p})$.

The Boltzmann equation for the gluon plasma \cite%
{Heinz:1984yq,Elze:1986qd,Biro:1993qt,Blaizot:1999xk,Baier:2000sb,Wang:2001dm}
reads 
\begin{eqnarray}
\frac{p^{\mu }}{E_{p}}\partial _{\mu }f_{p} &=&\frac{1}{N_{g}}\sum_{(n,l)}%
\frac{1}{N(n,l)}\int_{1\cdots \left( n-1\right) }d\Gamma _{1\cdots
l\rightarrow (l+1)\cdots (n-1)p}  \notag \\
&&\times \left[ (1+f_{p})\prod_{r=1}^{l}f_{r}%
\prod_{s=l+1}^{n-1}(1+f_{s})-f_{p}\prod_{r=1}^{l}(1+f_{r})%
\prod_{s=l+1}^{n-1}f_{s}\right] .  \label{eq:be1}
\end{eqnarray}%
The collision kernel 
\begin{equation}
d\Gamma _{1\cdots l\rightarrow (l+1)\cdots (n-1)p}\equiv \prod_{j=1}^{n-1}%
\frac{d^{3}\mathbf{p}_{j}}{(2\pi )^{3}2E_{j}}\frac{1}{2E_{p}}|M_{1\cdots
l\rightarrow (l+1)\cdots (n-1)p}|^{2}(2\pi )^{4}\delta
^{4}(\sum_{r=1}^{l}p_{r}-\sum_{s=l+1}^{n-1}p_{s}-p)  \label{eq:rate}
\end{equation}%
has summed over all colors and helicities of the initial and final states in
the matrix element squared. $N_{g}=2(N_{c}^{2}-1)=16$ is the color ($N_{c}=3$%
) and helicity degeneracy of a gluon. The $i$-th gluon is labeled as $i$
while the $n$-th gluon is labeled as $p$. For a process with $l$ initial and 
$(n-l)$ final gluons, the symmetry factor $N(n,l)=l!(n-l-1)!$. For example,
processes $12\rightarrow 3p$, $12\rightarrow 34p$, $123\rightarrow 4p$ yield 
$(n,l)=(4,2),(5,2),(5,3)$ and $N(n,l)=2,4,6$, respectively. $|M_{1\cdots
l\rightarrow (l+1)\cdots (n-1)p}|^{2}$ is the matrix element squared for the
process $1\cdots l\rightarrow (l+1)\cdots (n-1)p$ without average over the
degrees of freedom for incident gluons, i.e. it includes a factor $N_{g}^{2}$.

In vacuum, the matrix element squared for the 22 process is 
\begin{equation}
|M_{12\rightarrow 34}|^{2}=
8N_{g}(4\pi \alpha _{s}N_{c})^{2}
\left( 3-\frac{tu%
}{s^{2}}-\frac{su}{t^{2}}-\frac{st}{u^{2}}\right) ,  \label{eq:matrix-e22}
\end{equation}%
where $\alpha _{s}=g^{2}/(4\pi )$ is the strong coupling constant, and $%
(s,t,u)$ are the Mandelstam variables $s=(p_{1}+p_{2})^{2}$, $%
t=(p_{1}-p_{3})^{2}$ and $u=(p_{1}-p_{4})^{2}$.

For the 23 process \cite{Berends:1981rb,Ellis:1985er,Gottschalk:1979wq},
under the convention $\sum_{i=1}^{5}p_{i}=0$, we have 
\begin{eqnarray}
\left\vert M_{12345\rightarrow 0}\right\vert ^{2} &=&\left\vert
M_{0\rightarrow 12345}\right\vert ^{2}  \notag \\
&=&
\frac{1}{10}N_{g}(4\pi \alpha _{s}N_{c})^{3}
\left[ \left( 12\right)
^{4}+\left( 13\right) ^{4}+\left( 14\right) ^{4}+\left( 15\right)
^{4}+\left( 23\right) ^{4}\right.  \notag \\
&&\left. +\left( 24\right) ^{4}+\left( 25\right) ^{4}+\left( 34\right)
^{4}+\left( 35\right) ^{4}+\left( 45\right) ^{4}\right]  \notag \\
&&\times \sum\limits_{\mathrm{perm}\left\{ 1,2,3,4,5\right\} }\frac{1}{%
\left( 12\right) \left( 23\right) \left( 34\right) \left( 45\right) \left(
51\right) },  \label{eq:m3}
\end{eqnarray}%
where $(ij)\equiv p_{i}\cdot p_{j}$ and the sum is over all permutations of $%
\{1,2,3,4,5\}$. To convert to the convention $p_{1}+p_{2}=p_{3}+p_{4}+p_{5}$%
, we just perform the replacement: 
\begin{eqnarray}
\left\vert M_{12\rightarrow 345}\right\vert ^{2} &=&\left. \left\vert
M_{0\rightarrow 12345}\right\vert ^{2}\right\vert _{p_{1}\rightarrow
-p_{1},p_{2}\rightarrow -p_{2}},  \notag \\
\left\vert M_{345\rightarrow 12}\right\vert ^{2} &=&\left. \left\vert
M_{12345\rightarrow 0}\right\vert ^{2}\right\vert _{p_{1}\rightarrow
-p_{1},p_{2}\rightarrow -p_{2}}.  \label{m4}
\end{eqnarray}

In the medium, the gluon thermal mass serves as the infrared (IR) cut-off to
regularize IR sensitive observables. The most singular part of Eq.(\ref%
{eq:matrix-e22}) comes from the collinear region (i.e. either $t\approx 0$
or $u\approx 0$) which can be regularized by the HTL corrections to the
gluon propagators \cite{Weldon:1982aq,Pisarski:1988vd} and yields \cite%
{Heiselberg:1996xg}, 
\begin{equation}
|M_{12\rightarrow 34}|^{2}\approx 
4(4\pi \alpha _{s}N_{c})^{2}N_{g}
(4E_{1}E_{2})^{2}\left\vert \frac{1}{\mathbf{q}^{2}+\Pi
_{L}}-\frac{(1-\overline{x}^{2})\cos \phi }{\mathbf{q}^{2}(1-\overline{x}%
^{2})+\Pi _{T}}\right\vert ^{2},  \label{eq:htl}
\end{equation}%
where $q=p_{2}-p_{4}=(q_{0},\mathbf{q}),$ $\overline{x}=q_{0}/|\mathbf{q}|$
and $\phi $ is the angle between $\hat{\mathbf{p}}_{1}\times \hat{\mathbf{q}}
$ and $\hat{\mathbf{p}}_{2}\times \hat{\mathbf{q}}$. The HTL self-energies $%
\Pi _{L}$ (longitudinal) and $\Pi _{T}$ (transverse) are given by 
\begin{eqnarray}
\Pi _{L} &=&m_{D}^{2}\left[ 1-\frac{\overline{x}}{2}\ln \frac{1+\overline{x}%
}{1-\overline{x}}+i\frac{\pi }{2}\overline{x}\right] ,  \notag \\
\Pi _{T} &=&m_{D}^{2}\left[ \frac{\overline{x}^{2}}{2}+\frac{\overline{x}}{4}%
(1-\overline{x}^{2})\ln \frac{1+\overline{x}}{1-\overline{x}}-i\frac{\pi }{4}%
\overline{x}(1-\overline{x}^{2})\right] .  \label{HTL}
\end{eqnarray}%
The external gluon mass $m_{\infty }$\ (i.e. the asymptotic mass) is the
mass for an on-shell transverse gluon. In both the HTL approximation and the
full one-loop result, $m_{\infty }^{2}=\Pi _{T}\left( \left\vert \overline{x}%
\right\vert =1\right) =m_{D}^{2}/2$, where $m_{D}=(4\pi \alpha
_{s}N_{c}/3)^{1/2}T$ is the Debye mass. \ 

Previous perturbative analyses showed that the most important plasma effects
are the thermal masses $\sim gT$\ acquired by the hard thermal particles 
\cite{Blaizot:2000fc,Andersen:2002ey,CaronHuot:2007gq}. So a simpler (though
less accurate) treatment for the regulator is to insert $m_{D}$ to the gluon
propagator such that 
\begin{equation}
|M_{12\rightarrow 34}|^{2}\approx 
8N_{g}(4\pi \alpha _{s}N_{c}s)^{2}
\left[ \frac{1}{(t-m_{D}^{2})^{2}}+\frac{1}{(u-m_{D}^{2})^{2}}\right] .  
\label{MD}
\end{equation}%
It can be shown easily that Eqs. (\ref{eq:htl}) and (\ref{MD}) coincide in
the center-of-mass (CM) frame in vacuum. This treatment was used in Refs. 
\cite{Xu:2007ns,Biro:1993qt,Chen:2009sm}.

Eq. (\ref{MD}) is often expressed in $\mathbf{q}_{T}$, the transverse
component of $\mathbf{q}$ with respect to $\mathbf{p}_{1}$, in the CM frame.
If we just include the final state phase space of the $t$-channel, near
forward angle scatterings ($\mathbf{q}^{2}\approx \mathbf{q}_{T}^{2}\approx
0 $), then the backward angle contribution from the $u$-channel can be
included by multiplying the prefactor by a factor 2 
\begin{equation}
|M_{12\rightarrow 34}|_{CM}^{2}\underset{\mathbf{q}^{2}
\approx \mathbf{q}_{T}^{2}\approx 0} \approx 
16N_{g}(4\pi \alpha _{s}N_{c})^{2}
\frac{s^{2}}{(\mathbf{q}_{T}^{2}+m_{D}^{2})^{2}}.  
\label{AAA}
\end{equation}%
But if one includes the whole phase space in the calculation, then the
factor 2 is not needed: 
\begin{equation}
|M_{12\rightarrow 34}|_{CM}^{2}\underset{\mathbf{q}_{T}^{2}\approx 0} \approx 
8N_{g}(4\pi \alpha _{s}N_{c})^{2}
\frac{s^{2}}{(\mathbf{q}_{T}^{2}+m_{D}^{2})^{2}}.  \label{BBB}
\end{equation}
Note that the constraint $\mathbf{q}^{2}\approx 0$ is removed because both
the near forward and backward scatterings have small $\mathbf{q}_{T}^{2}$
but only the near forward scatterings have small $\mathbf{q}^{2}$.

For the 23 process, because the matrix element is already quite complicated,
we will just take $m_{D}$ as the internal gluon mass as was done in the $%
\eta $ computation in Ref. \cite{Chen:2010xk} and then estimate the errors.
In the $\sum_{i=1}^{5}p_{i}=0$ convention, one can easily show that an
internal gluon will have a momentum of $\pm (p_{i}+p_{j})$ rather than $\pm
(p_{i}-p_{j})$. Therefore, the gluon propagator factors $(ij)$ in the
denominator of Eq. (\ref{eq:m3}), is replaced by  
\begin{eqnarray}
(ij) &=&\frac{1}{2}[(p_{i}+p_{j})^{2}-m_{D}^{2}]  \notag \\
&=&p_{i}\cdot p_{j}\ +\frac{2m_{\infty }^{2}-m_{D}^{2}}{2}  \notag \\
&=&p_{i}\cdot p_{j}\ . \label{(ij)}
\end{eqnarray}%
Accidentally, $(ij)=p_{i}\cdot p_{j}$ is still correct 
after we have used the asymptotic mass for the external gluon mass. 
Then one applies Eq. (\ref{m4}) for the Boltzmann equation. In the numerator, 
the $(ij)^{4}$ combination is set by $T$ and is $\mathcal{O}(T^{8})$. So we can
still apply the substitution of Eq.(\ref{(ij)}), even if the $(ij)$ factors
might not have the inverse propagator form. The error is 
$\sim m_{D}^{2}(ij)^{3}=\mathcal{O}(\alpha _{s}T^{8})$, which is higher order in 
$\alpha _{s}$.

It is instructive to show that Eqs. (\ref{eq:m3},\ref{m4}) and (\ref{(ij)})
give the correct soft bremsstrahlung limit. Using the light-cone variable 
\begin{eqnarray}
p &=&\left( p^{+},p^{-},\mathbf{p}_{T}\right)   \notag \\
&\equiv &\left( p_{0}+p_{3},p_{0}-p_{3},p_{1},p_{2}\right) ,
\label{light-cone}
\end{eqnarray}%
we can rewrite one momentum configuration in the CM frame in terms of 
$p,p^{\prime },q$ and $k$: $p_{1}=p$, $p_{2}=p^{\prime }$, $p_{3}=p+q-k$, 
$p_{4}=p^{\prime }-q$ and $p_{5}=k$, with 
\begin{eqnarray}
p &=&\left( \sqrt{s},m_{\infty }^{2}/\sqrt{s},0,0\right) ,  \notag \\
p^{\prime } &=&\left( m_{\infty }^{2}/\sqrt{s},\sqrt{s},0,0\right) ,  \notag
\\
k &=&\left( y\sqrt{s},\left( k_{T}^{2}+m_{\infty }^{2}\right) /y\sqrt{s}%
,k_{T},0\right) ,  \notag \\
q &=&\left( q^{+},q^{-},\mathbf{q}_{T}\right) .  \label{momenta-light}
\end{eqnarray}%
The on-shell condition $p_{3}^{2}=p_{4}^{2}=m_{\infty }^{2}$ yields 
\begin{eqnarray}
q^{+} &\simeq &-q_{T}^{2}/\sqrt{s},  \notag \\
q^{-} &\simeq &\frac{k_{T}^{2}+yq_{T}^{2}-2y\mathbf{k}_{T}\cdot \mathbf{q}%
_{T}+(1-y+y^{2})m_{\infty }^{2}}{y\left( 1-y\right) \sqrt{s}}.
\end{eqnarray}
Here $y=k^+/p^+ = k_T e^{z}/\sqrt{s}$ is the light-cone momentum fraction of
the bremsstrahlung gluon and $z$ is its rapidity. In the central rapidity for
the bremsstrahlung gluon, i.e. $z \sim 0$, $y$ is toward zero.
In this case $p_5=k$ is very small compared to $p_1$ and $p_2$.

Now, in the limit $s\rightarrow \infty $, $y\rightarrow 0$, while keeping $y%
\sqrt{s}$ fixed, we have 
\begin{eqnarray}
p &=&\left( \sqrt{s},0,0,0\right) ,  \notag \\
p^{\prime } &=&\left( 0,\sqrt{s},0,0\right) ,  \notag \\
k &=&\left( y\sqrt{s},\left( k_{T}^{2}+m_{\infty }^{2}\right) /y\sqrt{s}%
,k_{T},0\right) ,  \notag \\
q &=&\left( 0,\left( k_{T}^{2}+m_{\infty }^{2}\right) /y\sqrt{s},\mathbf{q}%
_{T}\right) .  \label{GB limit}
\end{eqnarray}%
In this limit, $p_{1-4}$ are hard (their three momenta are $O\left( \sqrt{s}%
\right) $) while $q(=p_{2}-p_{4}=-p_{1}+p_{3}+p_{5})$ and $p_{5}=k$ are soft
(their three momenta are much smaller than $\sqrt{s}$). In this particular
limit of the phase space, the matrix element becomes 
\begin{equation}
\left\vert M_{12\rightarrow 345}\right\vert _{CM}^{2}\approx 
32(4\pi \alpha _{s}N_{c})^{3}N_{g}
\frac{s^{2}}{\left( k_{T}^{2}+m_{\infty }^{2}\right)
\left( q_{T}^{2}+m_{D}^{2}\right) \left[ \left( \mathbf{k}_{T}-\mathbf{q}%
_{T}\right) ^{2}+m_{D}^{2}\right] }.  \label{reduce-GB}
\end{equation}%
where the prefactor is equivalent to $3456\pi ^{3}\alpha _{s}^{3}N_{g}^{2}$
when $N_{c}=3$. 
Note that there are 6 different permutations of $(p_{3},p_{4},p_{5})$ which 
give the same expression as Eq. (\ref{reduce-GB}) due to the permutation symmetry of
Eq. (\ref{eq:m3}). Those permutations are corresponding to different
symmetric diagrams, just as the two permutations of $(p_{3},p_{4})$ in 
Eq. (\ref{MD}) give the $t$- and $u$-channel diagrams by the crossing symmetry.
Analogous to Eqs. (\ref{AAA}) and (\ref{BBB}), if we only include the
constraint phase space of $(p_{3},p_{4},p_{5})$, then we need to multiple
Eq. (\ref{reduce-GB}) by a factor $6$ to take into account the permutations
of $(p_{3},p_{4},p_{5})$. But if we include all the phase space in the
calculation, then Eqs. (\ref{eq:m3}) and (\ref{m4}) have to be used. Any
additional symmetry factor will result in multiple counting. 

The ratio of Eq. (\ref{reduce-GB}) to Eq. (\ref{BBB}) reproduces the
Gunion-Bertsch (GB) formula \cite{Gunion:1981qs} after taking 
$m_{D},m_{\infty }\rightarrow 0$. One can find the derivation of the GB
formula from Eq. (\ref{eq:m3}) in Appendix \ref{app1}. One can also expand
the \textquotedblleft exact\textquotedblright\ matrix element in 
Eqs. (\ref{eq:m3},\ref{m4}) in terms of $t/s$\ to extend the GB formula 
\cite{Das:2010hs,Das:2010yi,Abir:2010kc,Bhattacharyya:2011vy}.

An intuitive explanation of the LPM effect was given in 
Ref.\ \cite{Gyulassy:1991xb}: for the soft bremsstrahlung gluon with transverse
momentum $k_{T}$, the mother gluon has a transverse momentum uncertainty 
$\sim k_{T}$\ and a size uncertainty $\sim 1/k_{T}$. It takes the
bremsstrahlung gluon the formation time $t\sim 1/\left( k_{T}v_{T}\right)
\sim E_{k}/k_{T}{}^{2}$\ to fly far enough from the mother gluon to be
resolved as a radiation. But if the formation time is longer than the mean
free path $l_{mfp}\approx O(\alpha _{s}^{-1})$, then the radiation is
incomplete and it would be resolved as $gg\rightarrow gg$\ instead of 
$gg\rightarrow ggg$. Thus, the resolution scale is set by $t\leq l_{mfp}$.
This yields an IR cut-off $k_{T}^{2}\geq E_{k}/l_{mfp}\approx O(\alpha _{s})$ 
on the phase space \cite{Wang:1994fx}. Thus, the LPM effect reduces the 23
collision rate and will increase $\eta $\ and $\zeta $. Our previous
calculation on $\eta $\ using the Gunion-Bertsch formula shows that
implementing the $m_{D}$\ regulator gives a very close result to the LPM
effect \cite{Chen:2009sm}. Thus, we will estimate the size of the LPM effect
by increasing the external gluon mass $m_{g}$\ from $m_{\infty }$\ to $m_{D}$.

\section{An algorithm beyond variation to solve for $\zeta$}

Following the derivation of Ref. \cite{Jeon:1995zm}, the energy momentum
tensor of the weakly interacting gluon plasma in kinetic theory can be
modified as 
\begin{equation}
T_{\mu \nu }(x)=N_{g}\int \frac{\mathrm{d}^{3}\mathbf{p}}{(2\pi )^{3}E_{p}} 
f_{p}(x)\left( p_{\mu }p_{\nu }-\Sigma  (x ) g_{\mu \nu }\right) \ ,
\label{2}
\end{equation}
where $\Sigma (x)$ is an effective mass squared from 
the self-energy which encodes medium effects and 
$E_{p}=\sqrt{\mathbf{p}^{2}+m_{\infty }^{2}}$. 
When the system deviates from thermal equilibrium infinitesimally, $f_{p}(x)$ deviates from
its equilibrium value $f_{p}^{eq}=(e^{v\cdot p/T}-1)^{-1}$%
\begin{equation}
f_{p}=f_{p}^{eq}+\delta f_{p}.
\end{equation}%
And so does $T_{\mu \nu }$: 
\begin{equation}
\delta T_{\mu \nu }=N_{g}\int \frac{\mathrm{d}^{3}\mathbf{p}}{(2\pi
)^{3}E_{p}}\delta f_{p}\left( p_{\mu }p_{\nu }-v_{\mu }v_{\nu }T^{2}\frac{%
\partial m_{\infty }^{2}}{\partial T^{2}}\right) \ ,  \label{Yo-Lei}
\end{equation}%
where the energy momentum conservation $\partial ^{\mu }T_{\mu \nu }=0$ has
been imposed.

In hydrodynamics, small deviations from thermal equilibrium can be
systematically described by derivative expansions of hydrodynamical
variables with respect to spacetime. We will be working at the $\mathbf{v}%
(x)=0$ frame for a specific spacetime point $x$ (i.e. the local fluid rest
frame). This implies $\partial _{\nu }v^{0}=0$ after taking a derivative on $%
v_{\mu }(x)v^{\mu }(x)=1$. Then energy momentum conservation and thermal
dynamic relations (we have used the property that there is no conserved
charge in the system) in equilibrium allow us to express the time
derivatives $\partial _{t}T$ and $\partial _{t}\mathbf{v}$ in terms of the
spacial derivatives $\mathbf{\nabla }\cdot \mathbf{v}$ and $\mathbf{\nabla }%
T $. Thus, to the first derivative expansion of the hydrodynamical variables 
$\mathbf{v}$ and $T$, the bulk and shear viscosities are defined by the
small deviation of $T_{\mu \nu }$\ away from equilibrium:%
\begin{equation}
\delta T_{ij}=-\zeta \delta _{ij}\mathbf{\nabla }\cdot \mathbf{v}-\eta
\left( \frac{\partial v^{i}}{\partial x^{j}}+\frac{\partial v^{j}}{\partial
x^{i}}-\frac{2}{3}\delta _{ij}\mathbf{\nabla }\cdot \mathbf{v}\right) \ ,
\label{Yo-Hao}
\end{equation}%
\ where $i$ and $j$ are spacial indexes. Also, $\delta T_{0i}(x)=0$, since
the momentum density at point $x$ is zero in the local fluid rest frame, and
one defines $T_{00}$ to be the energy density in this frame. Therefore, 
\begin{equation}
\delta T_{00}=0\ =N_{g}\int \frac{\mathrm{d}^{3}\mathbf{p}}{(2\pi )^{3}E_{p}}%
\delta f_{p}\left( \mathbf{p}^{2}+\widetilde{m}^{2}\right) ,  \label{T00x}
\end{equation}%
where 
\begin{equation}
\widetilde{m}^{2}\equiv m_{\infty }^{2}-T^{2}\frac{\partial m_{\infty }^{2}}{%
\partial T^{2}}=-\frac{1}{6}N_{c}\beta (g^{2})T^{2}=\frac{11}{18}%
N_{c}^{2}\alpha _{s}^{2}T^{2}.
\end{equation}

Matching kinetic theory (Eq.(\ref{Yo-Lei})) to hydrodynamics (Eq.(\ref%
{Yo-Hao})) to the first derivative order, $\delta f_{p}$ can be parameterized
as 
\begin{equation}
\delta f_{p}=-\chi _{p}f_{p}^{eq}(1+f_{p}^{eq}),
\end{equation}%
where 
\begin{equation}
\chi _{p}=\frac{A(p)}{T}\mathbf{\nabla }\cdot \mathbf{v}+\frac{B_{ij}(p)}{T}%
\frac{1}{2}\left( \frac{\partial v^{i}}{\partial x^{j}}+\frac{\partial v^{j}%
}{\partial x^{i}}-\frac{2}{3}\delta _{ij}\mathbf{\nabla }\cdot \mathbf{v}%
\right) .  \label{eq:chi}
\end{equation}%
We can further write $B_{ij}(p)=B(p)(\mathbf{\hat{p}}_{i}
\mathbf{\hat{p}}_{j}-\frac{1}{3}\delta _{ij})$ with $\mathbf{\hat{p}}$ the unit vector in
the $\mathbf{p}$ direction. $A(p)$ and $B(p)$ are functions of $\mathbf{p}$.
They can be determined by the Boltzmann equation to give the solution of the
bulk and shear viscosities, respectively. In this work, we will focus on
solving the bulk viscosity.

Working to the first derivative order, the Boltzmann equation becomes a
linear equation in $\delta f_{p}$ which yields 
\begin{eqnarray}
\frac{\mathbf{p}^{2}}{3}-c_{s}^{2}(\mathbf{p}^{2}+\widetilde{m}^{2}) &=&%
\frac{E_{p}}{2N_{g}}\int_{123}d\Gamma _{12\rightarrow
3p}f_{1}^{eq}f_{2}^{eq}(1+f_{3}^{eq})(f_{p}^{eq})^{-1}[A_{3}+A_{p}-A_{1}-A_{2}]
\notag \\
&&+\frac{E_{p}}{4N_{g}}\int_{1234}d\Gamma _{12\rightarrow
34p}f_{1}^{eq}f_{2}^{eq}(1+f_{3}^{eq})(1+f_{4}^{eq})(f_{p}^{eq})^{-1}[A_{3}+A_{4}+A_{p}-A_{1}-A_{2}]
\notag \\
&&+\frac{E_{p}}{6N_{g}}\int_{1234}d\Gamma _{123\rightarrow
4p}f_{1}^{eq}f_{2}^{eq}f_{3}^{eq}(1+f_{4}^{eq})(f_{p}^{eq})^{-1}[A_{4}+A_{p}-A_{1}-A_{2}-A_{3}].
\label{eq:p3}
\end{eqnarray}%
Here we have used the notation $A_{p}\equiv A(p)$ and $A_{i}\equiv A(p_{i})$
with $i=1,2,3,4$. The speed of sound squared $c_{s}^{2}$ is defined as \cite%
{Jeon:1995zm,Arnold:2006fz}, 
\begin{equation}
c_{s}^{2}\equiv \frac{\partial P}{\partial \epsilon }=\frac{\int
d^{3}pf_{p}^{eq}(1+f_{p}^{eq})\mathbf{p}^{2}}{3\int
d^{3}pf_{p}^{eq}(1+f_{p}^{eq})(\mathbf{p}^{2}+\widetilde{m}^{2})}.
\end{equation}%
Then Eqs. (\ref{Yo-Lei},\ref{Yo-Hao},\ref{eq:chi}) yield 
\begin{equation}
\zeta =\frac{N_{g}}{T}\int \frac{d^{3}p}{(2\pi )^{3}E_{p}}%
f_{p}^{eq}(1+f_{p}^{eq})\left[ \frac{1}{3}\mathbf{p}^{2}-c_{s}^{2}(\mathbf{p}%
^{2}+\widetilde{m}^{2})\right] A(p),  \label{eq:bulk1}
\end{equation}%
where we have added the $c_{s}^{2}$ term which is proportional to $\delta
T_{00}=0$ for convenience. By substituting Eq. (\ref{eq:p3}) into Eq. (\ref%
{eq:bulk1}), we obtain 
\begin{eqnarray}
\zeta &=&\frac{1}{8T}\int \prod_{i=1}^{4}\frac{d^{3}k_{i}}{(2\pi )^{3}2E_{i}}%
|M_{12\rightarrow 34}|^{2}(2\pi )^{4}\delta ^{4}(E_{1}+E_{2}-E_{3}-E_{4}) 
\notag \\
&&\times
(1+f_{1}^{eq})(1+f_{2}^{eq})f_{3}^{eq}f_{4}^{eq}[A_{3}+A_{4}-A_{1}-A_{2}]^{2}
\notag \\
&&+\frac{1}{12T}\int \prod_{i=1}^{5}\frac{d^{3}k_{i}}{(2\pi )^{3}2E_{i}}%
|M_{12\rightarrow 345}|^{2}(2\pi )^{4}\delta
^{4}(E_{1}+E_{2}-E_{3}-E_{4}-E_{5})  \notag \\
&&\times
(1+f_{1}^{eq})(1+f_{2}^{eq})f_{3}^{eq}f_{4}^{eq}f_{5}^{eq}[A_{3}+A_{4}+A_{5}-A_{1}-A_{2}]^{2}.
\label{eq:bulk3}
\end{eqnarray}%
By the definition of $c_{s}$, the following integral vanishes: 
\begin{equation}
\int \frac{d^{3}p}{(2\pi )^{3}E_{p}}f_{p}^{eq}(1+f_{p}^{eq})\left[ \frac{%
\mathbf{p}^{2}}{3}-c_{s}^{2}(\mathbf{p}^{2}+\widetilde{m}^{2})\right]
E_{p}=0.  \label{eq:zero-energy}
\end{equation}%
We will use this property later.

Now we first review the arguments that cast the computation of $\zeta $ as a
variational problem \cite{Resibois,Arnold:2003zc}. Then we show how one can
go beyond variation to find the answer systematically. Let us rewrite Eq.(%
\ref{eq:p3}) schematically as%
\begin{equation}
\left\vert S\right\rangle =C\left\vert A\right\rangle ,  \label{A}
\end{equation}%
and Eqs. (\ref{eq:bulk1},\ref{eq:bulk3}) as 
\begin{equation}
\zeta =\left\langle A|S\right\rangle =\left\langle A\left\vert C\right\vert
A\right\rangle .  \label{B}
\end{equation}%
Note that Eq. (\ref{B}) is just a projection of Eq. (\ref{A}). Using $%
\left\vert A\right\rangle =C^{-1}\left\vert S\right\rangle $ from Eq. (\ref%
{A}),%
\begin{equation}
\zeta =\left\langle S\left\vert C^{-1}\right\vert S\right\rangle .  \label{C}
\end{equation}

Technically, finding an ansatz $A_{anz}$ that satisfies the projected equation $%
\left\langle S|A_{anz}\right\rangle =\left\langle A_{anz}\left\vert
C\right\vert A_{anz}\right\rangle $ of (\ref{B}) is easier than solving the
original integral equation (\ref{A}). But this will not give the correct
viscosity if $C\left\vert A_{anz}\right\rangle \neq \left\vert
S\right\rangle $. However, the resulting bulk viscosity is always 
less than the real one,
\begin{eqnarray}
\zeta _{anz} &=&-\left\langle A_{anz}\left\vert C\right\vert
A_{anz}\right\rangle +2\left\langle A_{anz}|S\right\rangle  \notag \\
&=&-\left\langle A_{anz}^{\prime }\left\vert C\right\vert A_{anz}^{\prime
}\right\rangle +\left\langle S\left\vert C^{-1}\right\vert S\right\rangle 
\notag \\
&\leq &\left\langle S\left\vert C^{-1}\right\vert S\right\rangle =\zeta ,
\label{bbb}
\end{eqnarray}%
where $\left\vert A_{anz}^{\prime }\right\rangle \equiv \left\vert
A_{anz}\right\rangle -C^{-1}\left\vert S\right\rangle $ and $\left\langle
A_{anz}^{\prime }\left\vert C\right\vert A_{anz}^{\prime }\right\rangle $ is
real and non-negative. Thus, a variational calculation of $\zeta $ is
possible: one just demands $\left\langle S|A_{anz}\right\rangle
=\left\langle A_{anz}\left\vert C\right\vert A_{anz}\right\rangle $ and try
to find the maximum $\zeta _{anz}$. In what follows, we show an algorithm
(see Eqs.(\ref{aaa})-(\ref{zzz})) that will approach the maximum $\zeta
_{anz}$ systematically.

We will choose a basis $\left\{ \tilde{A}_{i}|i=1,2,\ldots ,n\right\} $ with 
$n$ orthonormal functions satisfying 
\begin{equation}
\left\langle \tilde{A}_{i}\left\vert C\right\vert \tilde{A}_{j}\right\rangle
=\delta _{ij}\ .  \label{aaa}
\end{equation} 
We impose the following condition for $\tilde{A}_{i}$
\begin{equation}
\int \frac{d^{3}p}{(2\pi )^{3}E_{p}}f_{p}^{eq}(1+f_{p}^{eq})(\mathbf{p}^{2}+%
\widetilde{m}^{2})\tilde{A}_{i}(p)=0,  \label{AA}
\end{equation} 
and we take the ansatz for $A$ 
\begin{equation}
A_{anz}^{(n)}\equiv \sum_{i=1}^{n}d_{i}\tilde{A}_{i},
\end{equation}
so that the constraint $\delta T_{00}=0$ is automatically satisfied. 
Then Eq. (\ref{B}) yields 
\begin{equation}
\zeta _{anz}^{(n)}=\sum_{i=1}^{n}d_{i}\left\langle \tilde{A}%
_{i}|S\right\rangle =\sum_{ij=1}^{n}d_{i}d_{j}\left\langle \tilde{A}%
_{i}\left\vert C\right\vert \tilde{A}_{j}\right\rangle
=\sum_{i=1}^{n}d_{i}^{2}.  \label{D}
\end{equation}%
This equation does not determine $d_{i}$ uniquely. However, what we want is
the solution that maximizes $\zeta _{anz}^{(n)}$, which is unique.
It can be computed by rewriting Eq.(\ref{D}) as 
\begin{eqnarray}
\zeta _{anz}^{(n)} &=&\sum_{i=1}^{n}\left( 2d_{i}\left\langle \tilde{A}%
_{i}|S\right\rangle -d_{i}^{2}\right)   \notag \\
&=&\sum_{i=1}^{n}\left\langle \tilde{A}_{i}|S\right\rangle
^{2}-\sum_{i=1}^{n}\left( d_{i}-\left\langle \tilde{A}_{i}|S\right\rangle
\right) ^{2}.  \label{E}
\end{eqnarray}%
Then the solution 
\begin{equation}
d_{i}=\left\langle \tilde{A}_{i}|S\right\rangle   \label{ddd}
\end{equation}%
satisfies the projected equation (\ref{D}). It is also the solution we are
looking for which maximizes $\zeta _{anz}^{(n)}$. This solution yields%
\begin{equation}
\zeta _{anz}^{(n)}=\sum_{i=1}^{n}\left\langle \tilde{A}_{i}|S\right\rangle
^{2}.
\end{equation}%
Since $\left\langle \tilde{A}_{i}|S\right\rangle $ is real, $\zeta
_{anz}^{(n)}$ is monotonically increasing with respect to $n$. Also, we have 
$\zeta \geq \zeta _{anz}$ from Eq.\ (\ref{bbb}). This yields%
\begin{equation}
\zeta _{anz}^{(n)}\leq \zeta _{anz}^{(n+1)}\leq \zeta _{anz}^{(n\rightarrow
\infty )}=\zeta ,  \label{zzz}
\end{equation}%
which means we can systematically approaches $\zeta $ from below by
increasing $n$, then we will see $\zeta _{anz}^{(n)}$ becomes larger and
larger. We stop at a finite $n$ when a good convergence of the series $\zeta
_{anz}^{(n)}$ is observed. So this algorithm systematically
approaches $\zeta $ from below.


We will use the following basis 
\begin{equation}
\tilde{A}_{i} =\sum_{j=0}^{i}c_{j}(E_{p}/T)^{j},  
\end{equation} 
where $A_{anz}^{(n)}$ is given by 
\begin{equation}
A_{anz}^{(n)} = \sum_{i=1}^{n}d_{i}\tilde{A}_{i}
=\sum_{i=0}^{n}\tilde{c}_{i}(E_{p}/T)^{i}. 
\label{eq:expansion-a}
\end{equation}
The orthonormal condition in Eq.\ (\ref{aaa}) determines $c_{j}$ and 
Eq.(\ref{ddd}) determines $d_{i}$. Equivalently, one can also solve for $\tilde{c}%
_{i}$ directly by demanding $\left\langle S|A_{anz}^{(n)}\right\rangle
=\left\langle A_{anz}^{(n)}\left\vert C\right\vert
A_{anz}^{(n)}\right\rangle $ is satisfied and the $\tilde{c}_{i}$ solution
gives the maximum $\zeta _{anz}^{(n)}$. Note that although the $E_{p}$ term
does not contribute to $\left\langle A_{anz}^{(n)}\left\vert C\right\vert
A_{anz}^{(n)}\right\rangle $ or $\left\langle S|A_{anz}^{(n)}\right\rangle $%
, it does not mean the $c_{1}$ coefficient is not fixed in this procedure. $%
c_{1}$ is fixed by the constraint $\delta T_{00}=0$.

An alternative basis is used in Ref. \cite{Arnold:2006fz}:%
\begin{equation}
A_{anz}^{\prime (n)}=\sum_{i=1}^{n}c_{i}^{\prime }\frac{(p/T)^{i}}{%
(p/T+1)^{n-2}}+d^{\prime }E_{p}.
\end{equation}%
The two bases give give consistent $\zeta $. For example, at $\alpha
_{s}=0.1 $, the agreement is better than $1\%$\ when we work up to $n=6$.

\section{N$_{c}$ Scaling and Numerical results}

\subsection{$N_{c}$ Scaling}

\label{Nc}

Viscosities of a general $SU(N_{c})$\ pure gauge theory can be obtained by
simply rescaling the $SU(3)$\ result. Using the above formulas, it is easy
to show that 
\begin{equation}
\zeta =N_{g}g_{1}\left( \alpha _{s}N_{c}\right) T^{3},\ \eta
=N_{g}g_{2}\left( \alpha _{s}N_{c}\right) T^{3},
\end{equation}%
where $g_{1}$\ and $g_{2}$\ are dimensionless functions of $\alpha _{s}N_{c}$%
\ only. This, together with $s\propto N_{g}$, yields 
\begin{equation}
\frac{\zeta }{s}=h_{1}\left( \alpha _{s}N_{c}\right) ,\ \frac{\eta }{s}%
=h_{2}\left( \alpha _{s}N_{c}\right) ,
\end{equation}%
where $h_{1}$\ and $h_{2}$\ are also dimensionless functions of $\alpha
_{s}N_{c}$\ only. Thus, our $\zeta /s$, $\eta /s$\ and $\zeta /\eta $\ vs. $%
\alpha _{s}N_{c}$\ curves in Fig. \ref{fig:Bulk2Shear} are universal and
suitable for a general $SU(N_{c})$\ pure gauge theory. From now on, $N_{c}=3$%
\ unless otherwise specified. One can always rescale the results to an
arbitrary $N_{c}$.

\subsection{Leading-Log result}

As discussed above, in the leading-log approximation, one just needs to
focus on the small $q_{T}$ contribution from the 22 process while setting $%
c_{0}=0$. Furthermore, it was shown in \cite{Baym:1990uj,Heiselberg:1994vy}
that using the HTL regulator (\ref{eq:htl}) gives the same LL shear
viscosity to that using the $m_{D}$ regulator (\ref{MD}). For the bulk
viscosity, this is also true. We obtained the same LL result as \cite%
{Arnold:2006fz}, 
\begin{equation}
\zeta _{LL}\simeq 0.44\frac{T^{3}\alpha _{s}^{2}}{\ln (1/g)}.  \label{LL}
\end{equation}%
This can be compared with \cite{Arnold:2000dr,Chen:2010xk}%
\begin{equation}
\eta _{LL}\simeq 0.17\frac{T^{3}}{\alpha _{s}^{2}\ln (1/g)}.
\end{equation}%
For a gluon plasma, we have 
\begin{equation}
\frac{\zeta _{LL}}{\eta _{LL}}\simeq 2.6\alpha _{s}^{4}=48\left(
1/3-c_{s}^{2}\right) ^{2}.  \label{LL-zeta-eta}
\end{equation}%
This is parametrically the same as $\zeta /\eta =15\left(
1/3-c_{s}^{2}\right) ^{2}$ for the absorption and emission of light quanta
(e.g. photons, gravitons or neutrinos) by the medium \cite{Weinberg:1971mx}.
In the $\alpha _{s}\ll 1$ region where QCD is perturbative, $\zeta \ll \eta $%
. Using the entropy density for non-interacting gluons, $s=N_{g}\frac{2\pi
^{2}}{45}T^{3}$, we have 
\begin{equation}
\frac{\zeta _{LL}}{s}\simeq 0.063\frac{\alpha _{s}^{2}}{\ln (1/g)},\ \ \frac{%
\eta _{LL}}{s}\simeq \frac{0.025}{\alpha _{s}^{2}\ln (1/g)}.
\end{equation}

\subsection{Numerical results of $\protect\eta $ and\ $\protect\zeta $}

\begin{figure}[tbp]
\caption{(Color online) (a) The ratio of our numerical result (denoted as
\textquotedblleft exact\textquotedblright ) for the shear viscosity 
$\eta $ to AMY's. The error is bounded by the upper and lower bound. The
result using the GB matrix element [Eq. (\ref{reduce-GB})] is also
shown. (b) Comparison of our result of $\eta /\eta _{22}$ 
with those of AMY, XG and \textquotedblleft GB$\times 6$%
\textquotedblright (see the text). For our result the HTL gluon propagator
is used for the 22 process, the \textquotedblleft exact\textquotedblright\
matrix element Eqs. (\ref{eq:m3},\ref{m4},\ref{(ij)}) 
are used for the 23 process, and the external gluon mass is set to 
$m_{\infty }$. In the left panel, the error bands are shown. AMY's result is
taken from Ref. \cite{Arnold:2000dr,Arnold:2003zc} and XG's 
is from Ref. \cite{Wesp:2011yy}. }
\label{fig:shear23}
\includegraphics[scale=0.4]{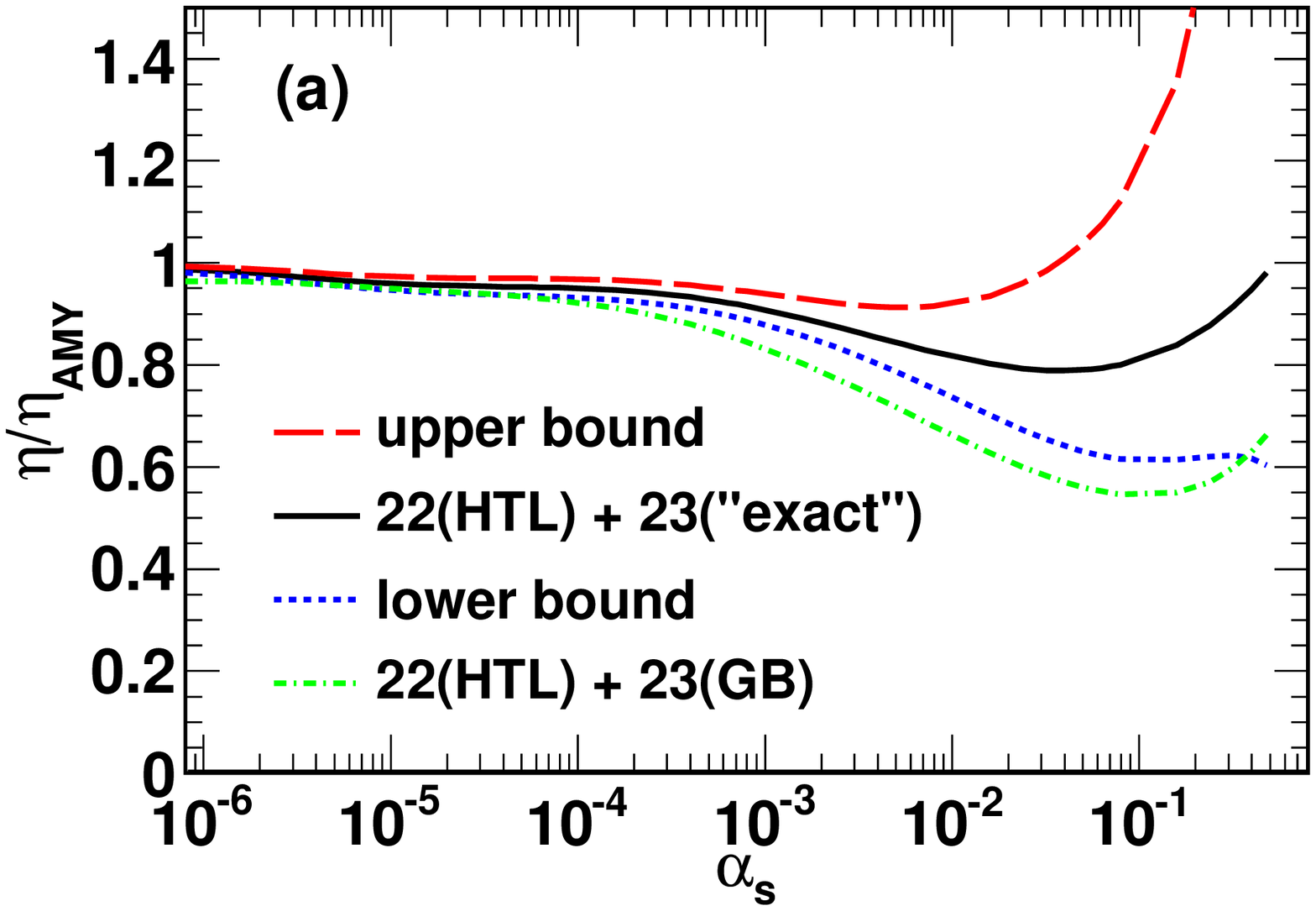} 
\includegraphics[scale=0.4]{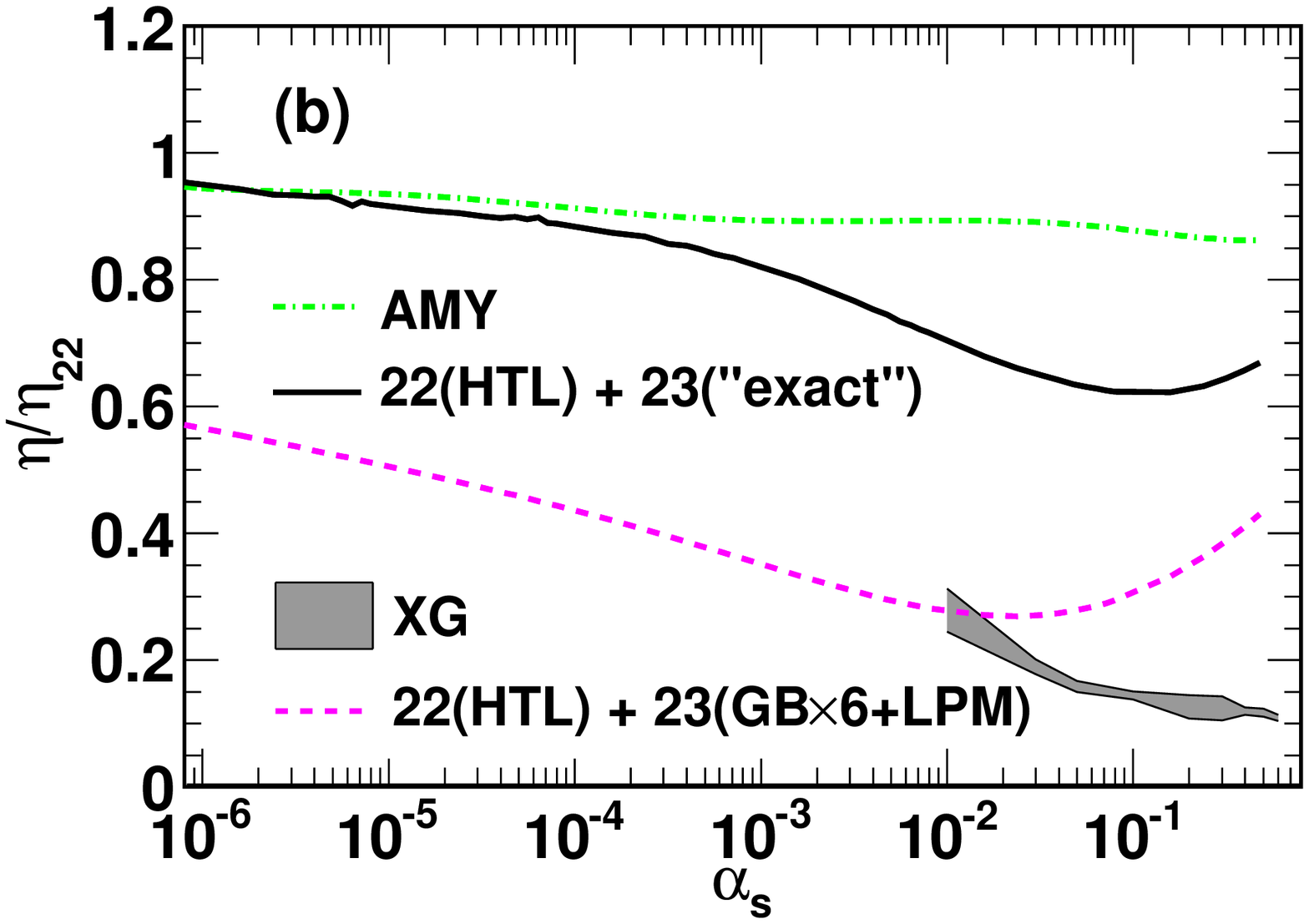}
\end{figure}

\begin{figure}[tbp]
\caption{(Color online) $\eta /s$ with various inputs.}
\label{comp-gb-ex}
\includegraphics[scale=0.4]{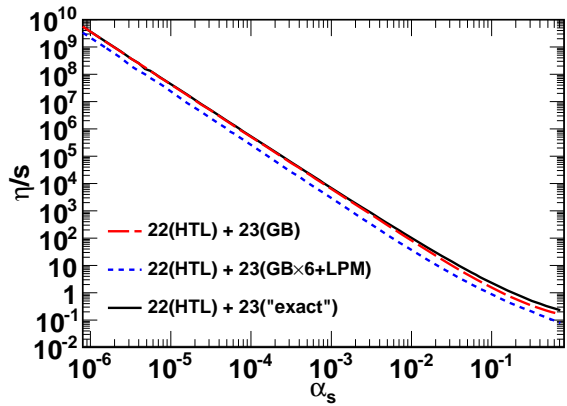}
\end{figure}

In our calculation, we use the HTL propagator for the 22 process. For the 23
process, for technical reasons, we use the internal gluon mass $m_{D}$
instead of the HTL propagator in Eqs. (\ref{eq:m3}-\ref{eq:htl},\ref{(ij)}), 
$E_{p}=\sqrt{\mathbf{p}^{2}+m_{\infty }^{2}}$ in kinematics and $f_{p}^{eq}$
for the external gluon distribution. The errors from not implementing HTL 
propagator in the 23 process and the modeling of the Landau-Pomeranchuk-Migdal (LPM) effect, 
and from the uncalculated $O(\sqrt{\alpha _{s}})$ higher order corrections are estimated
in Appendix \ref{app2}.

In Fig. \ref{fig:shear23}, we show our main result for the shear viscosity 
$\eta $ in our previous paper \cite{Chen:2009sm}, together with the
theoretical error band bounded by \textquotedblleft upper
bound\textquotedblright\ and \textquotedblleft lower
bound\textquotedblright\ curves. Note that previously we estimated the
higher order effect to be $O(\alpha _{s})$ suppressed. But since the
expansion parameter in finite temperature field theory is $g$ instead of 
$g^{2}$, we enlarge the error of the higher order effect to $O(\sqrt{\alpha
_{s}})$\ here. The result agrees with AMY's result within errors in the left
panel although our central value is lower at larger $\alpha _{s}$.\ If we
replace the \textquotedblleft exact\textquotedblright\ matrix element of
Eqs. (\ref{eq:m3}-\ref{eq:htl},\ref{(ij)}) by the GB matrix element of 
Eq. (\ref{reduce-GB}), then $\eta $ is reduced but still close to the estimated
lower bound. This means the 23 collision rate in GB is bigger than that in
\textquotedblleft exact.\textquotedblright 

The effect of the 23 process can be seen more clearly in the ratio $\eta
/\eta _{22}$ ($\eta _{22}$ means the shear viscosity with the 22 process
included only) shown in the right panel, where we also show AMY's and XG's
results for comparison. In AMY's result \cite{Arnold:2000dr,Arnold:2003zc},
the near collinear $1\leftrightarrow 2$ process gives $\eta /\eta _{22}$
close to unity. This implies their 12 collision is just a small perturbation
to the 22 rate. However, XG employ the soft gluon bremsstrahlung
approximation in the matrix element for the 23 process, gives $\eta /\eta
_{22}\simeq 0.11\sim 0.16$ around 1/8 in Ref. \cite{Xu:2007ns}, indicating
that their 23 collision rate is about 7 times the 22 one. In their improved
treatment using the Kubo relation \cite{Wesp:2011yy}, they give $\eta /\eta
_{22}\simeq 0.1\sim 0.3$, indicating that the 23 collision rate is about 2$%
\sim $9 times the 22 rate.

Our central result lies between AMY's and XG's results. However, even consider the
lower bound, our 23 rate does not get bigger than the 22 rate. Thus, it is
qualitatively consistent with AMY's result but inconsistent with XG's
result. When compared with AMY's result, in addition to the error band shown
in the left panel, there is still $\sim 10\%$\ difference at $\alpha
_{s}=0.01$. This is consistent with the $\sim m_{g}^{2}/T^{2}$\ effect from
using different inputs for external gluon mass--- we use $m_{g}$\ while AMY
use zero. 

We find that we cannot reproduce XG's result unless we use a 23 matrix
element squared at least 6 times larger. To compare with XG's calculation,
we use the same $m_{\infty }=0$\ and LPM effect as XG, and

\begin{equation}
\left\vert M_{12\rightarrow 345}\right\vert _{CM}^{2}\rightarrow 6\times
54g^{6}N_{g}^{2}\frac{q_{T}^{2}s^{2}}{k_{T}^{2}\left(
q_{T}^{2}+m_{D}^{2}\right) ^{2}\left[ \left( \mathbf{k}_{T}-\mathbf{q}%
_{T}\right) ^{2}+m_{D}^{2}\right] },  \label{GB6}
\end{equation}%
which is a slightly different variation of the GB matrix element squared of
Eq. (\ref{reduce-GB}) multiplied by a factor 6 (denoted as \textquotedblleft
GB$\times 6$\textquotedblright ). This reproduces XG's result at 
$\alpha _{s}=0.01$. The origin of this discrepancy is yet to be resolved. 

In Fig. \ref{comp-gb-ex}, $\eta /s$ with various inputs are shown. At $\alpha
_{s}=0.3$ and 0.6,\ the GB$\times 6$\ curve yields $\eta /s=0.19$ and
0.09, respectively, while XG has 0.13 and 0.08. The central value of the
\textquotedblleft exact\textquotedblright\ result is about two times lager. 

\begin{figure}[tbp]
\caption{(Color online) (a) Comparison of our result for the bulk viscosity 
$\zeta $ and its error band (see the appendix) with ADM's result.
`LL' denotes the leading-log result of Eq.(\ref{LL}). `ADM LO'
denotes ADM's leading order result read off from Ref.\ \cite%
{Arnold:2006fz} (only available for $\alpha _{s}\gtrsim 8\times
10^{-4}$). (b) Our full result for $\zeta $ (denoted as
22+23) and $\zeta $ with the 23 process only (denoted as 23). }
\label{fig:bulk23}
\includegraphics[scale=0.4]{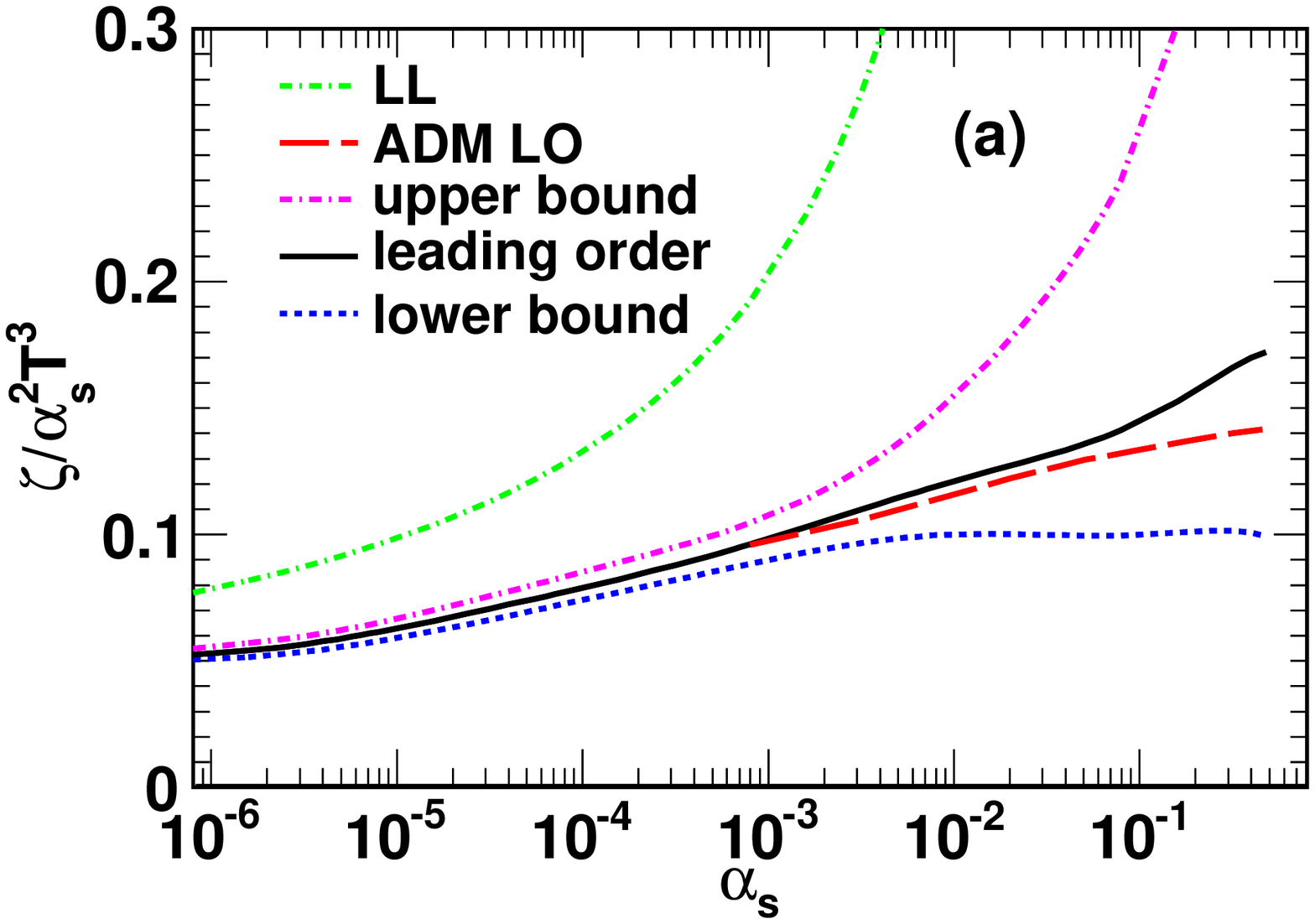} %
\includegraphics[scale=0.4]{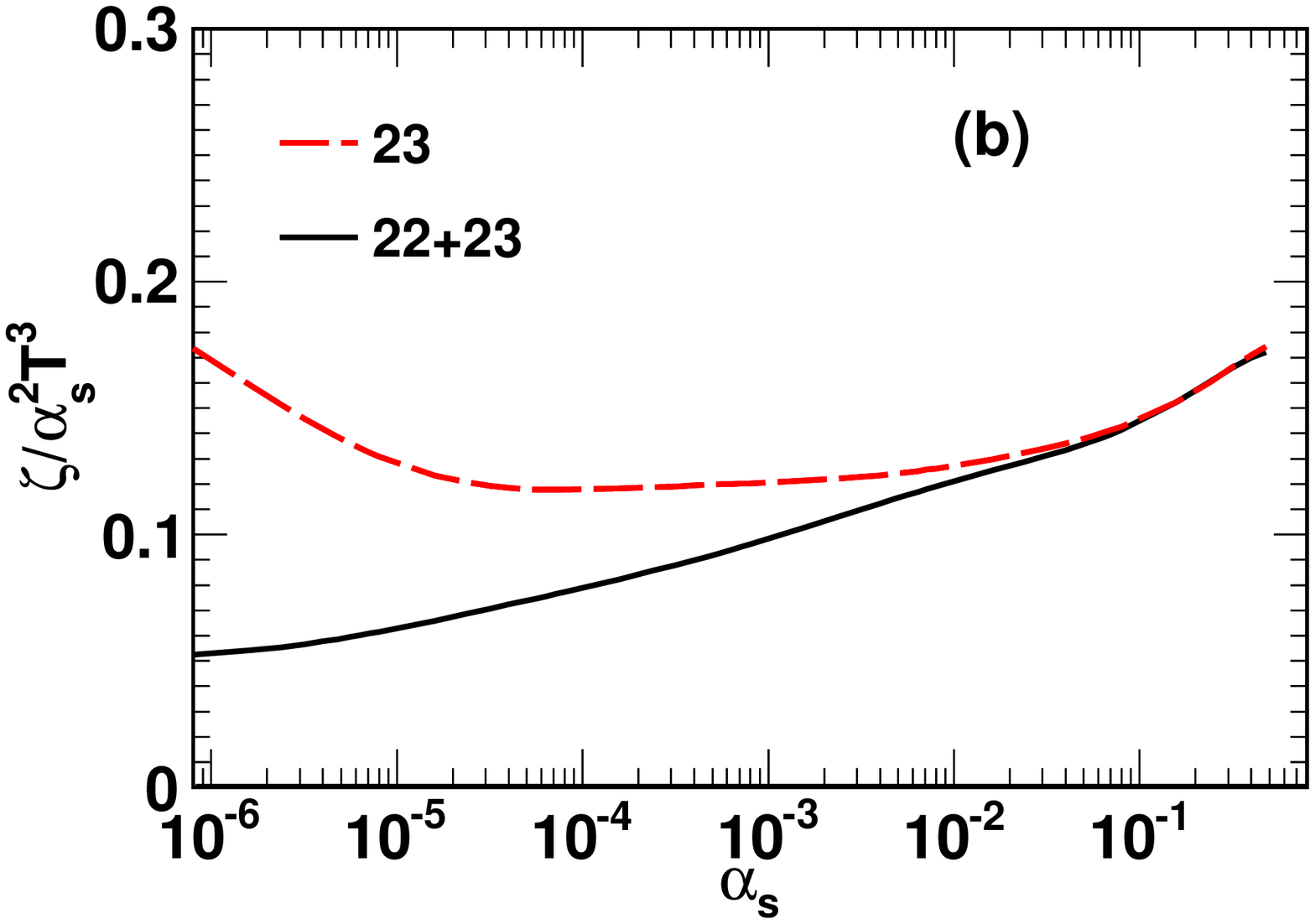}
\end{figure}

Our result for the bulk viscosity $\zeta $ using the \textquotedblleft
exact\textquotedblright\ matrix element for the 23 process is shown in Fig. %
\ref{fig:bulk23}. We have worked up to $n=6$ and seen good convergence. For
example, we obtain $\zeta _{anz}^{(n)}\sim $ 95\%, 98\%, 99.5\% of $\zeta
_{anz}^{(6)}$ at $\alpha _{s}=10^{-4}$ for $n=3,4,5$ respectively. The
convergence for larger $\alpha _{s}$ is even better. When $\alpha
_{s}\lesssim 10^{-8}$, our result approaches the LL one. At larger $\alpha
_{s}$, the 23 process becomes more important such that when $\alpha
_{s}\gtrsim 0.1$, $\zeta $ is saturated by the 23 contribution (see the
right panel of Fig. \ref{fig:bulk23}). Our result agrees with that of ADM 
\cite{Arnold:2006fz} in the full range of $\alpha _{s}$ within the error
band explained in the Appendix \ref{app2}.

\subsection{Angular Correlation in 23 Process}

\begin{figure}[tbp]
\caption{(Color online) The distribution of $\theta $ in the fluid local rest
frame. (a) Weighted by the phase space and the Bose-Einstein
distribution functions only. (b) Weighted by the contribution to 
$\eta _{23}$ with the \textquotedblleft exact\textquotedblright\
matrix element. (c) Weighted by the contribution to $\eta _{23}$ 
with the GB matrix element. 
The upper panel is normalized to unity for each coupling constant. The lower panel shows the  location of the peak $\theta^{peak}$, the average value $<\theta>$ and the variation $\sigma_{\theta}$ of the angle. 
The angle is in the unit of radian. }
\label{fig:theta-lrf}
\includegraphics[width=0.3\textwidth]{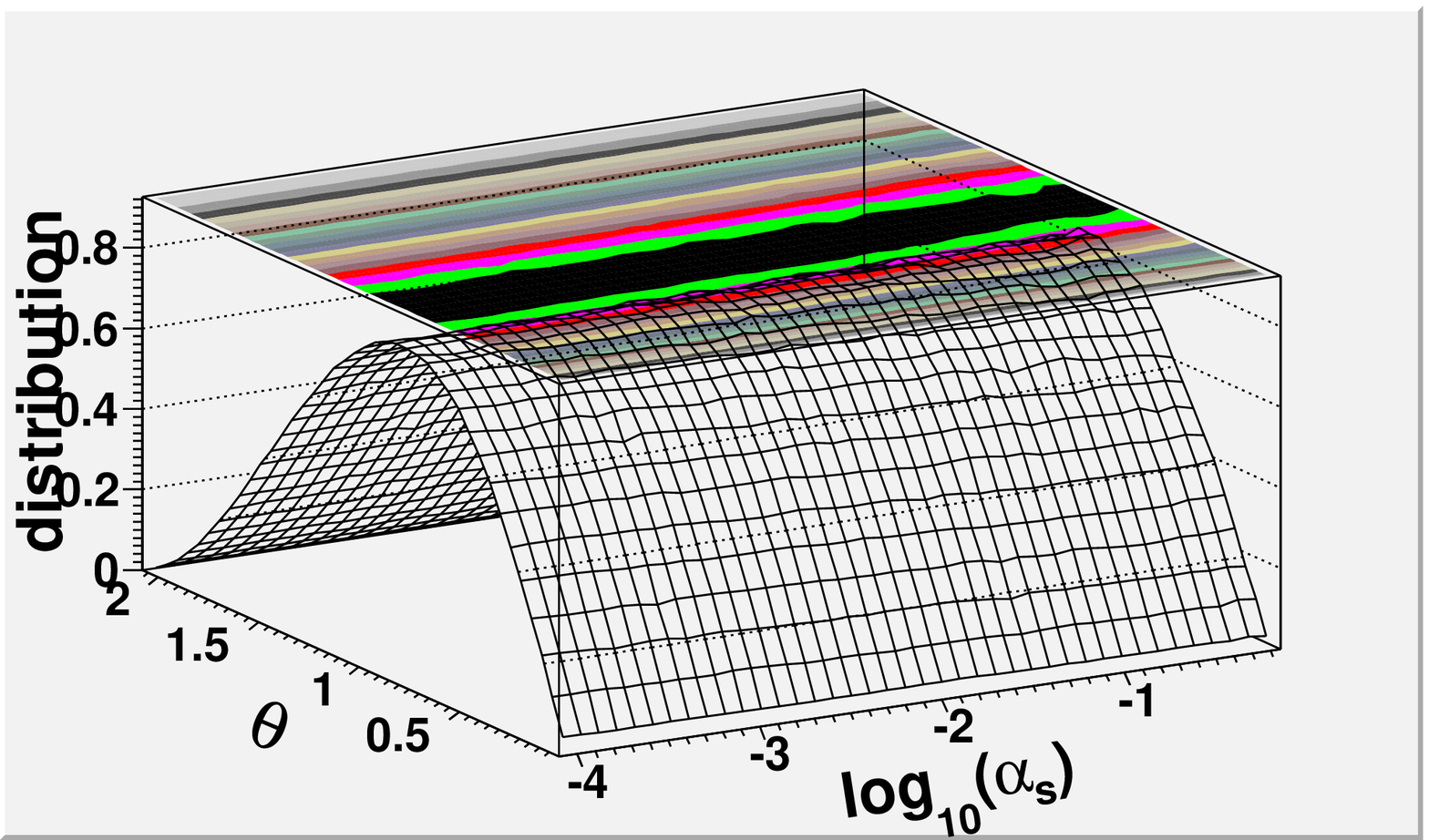} 
\includegraphics[width=0.3\textwidth]{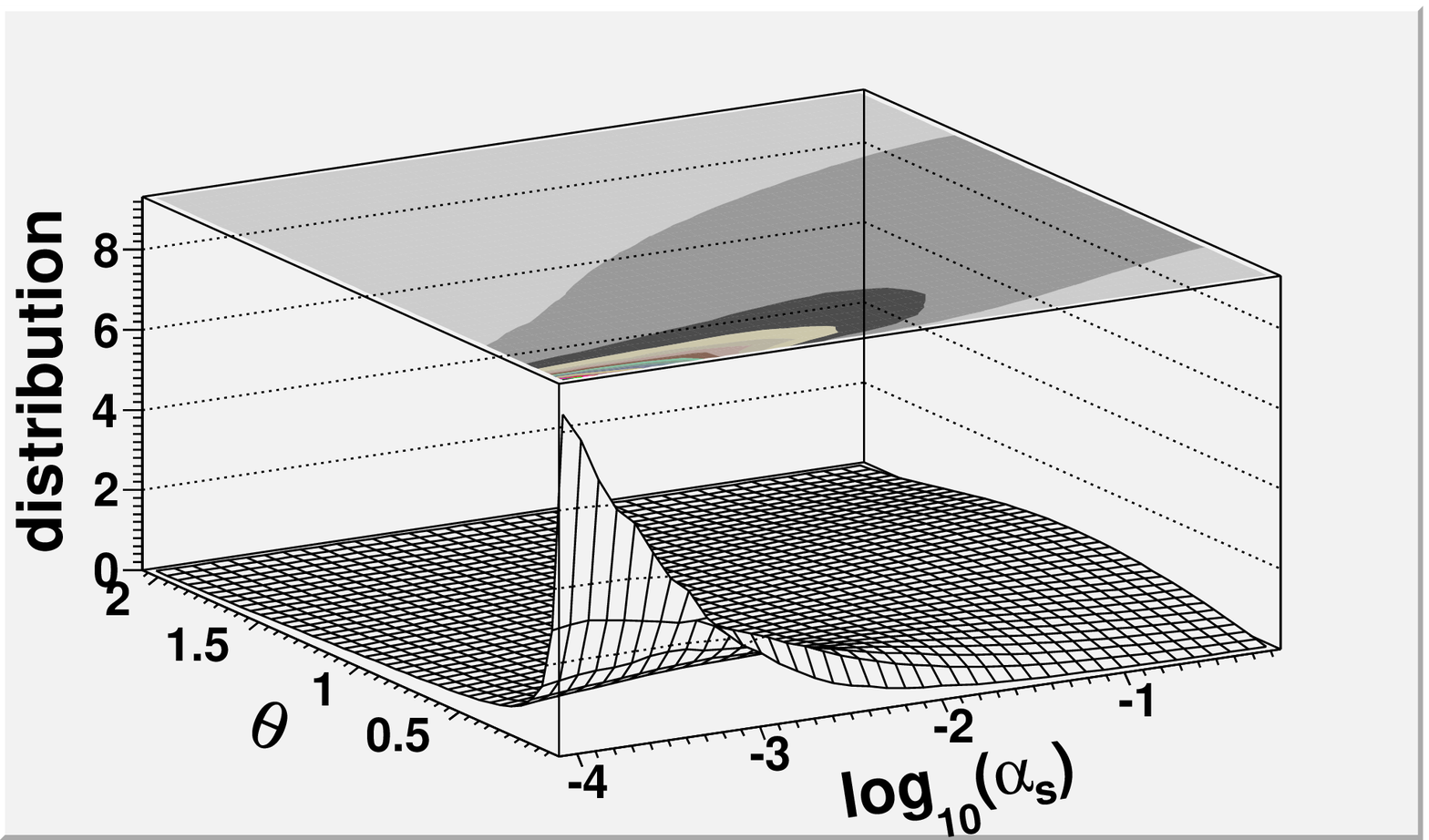} 
\includegraphics[width=0.3\textwidth]{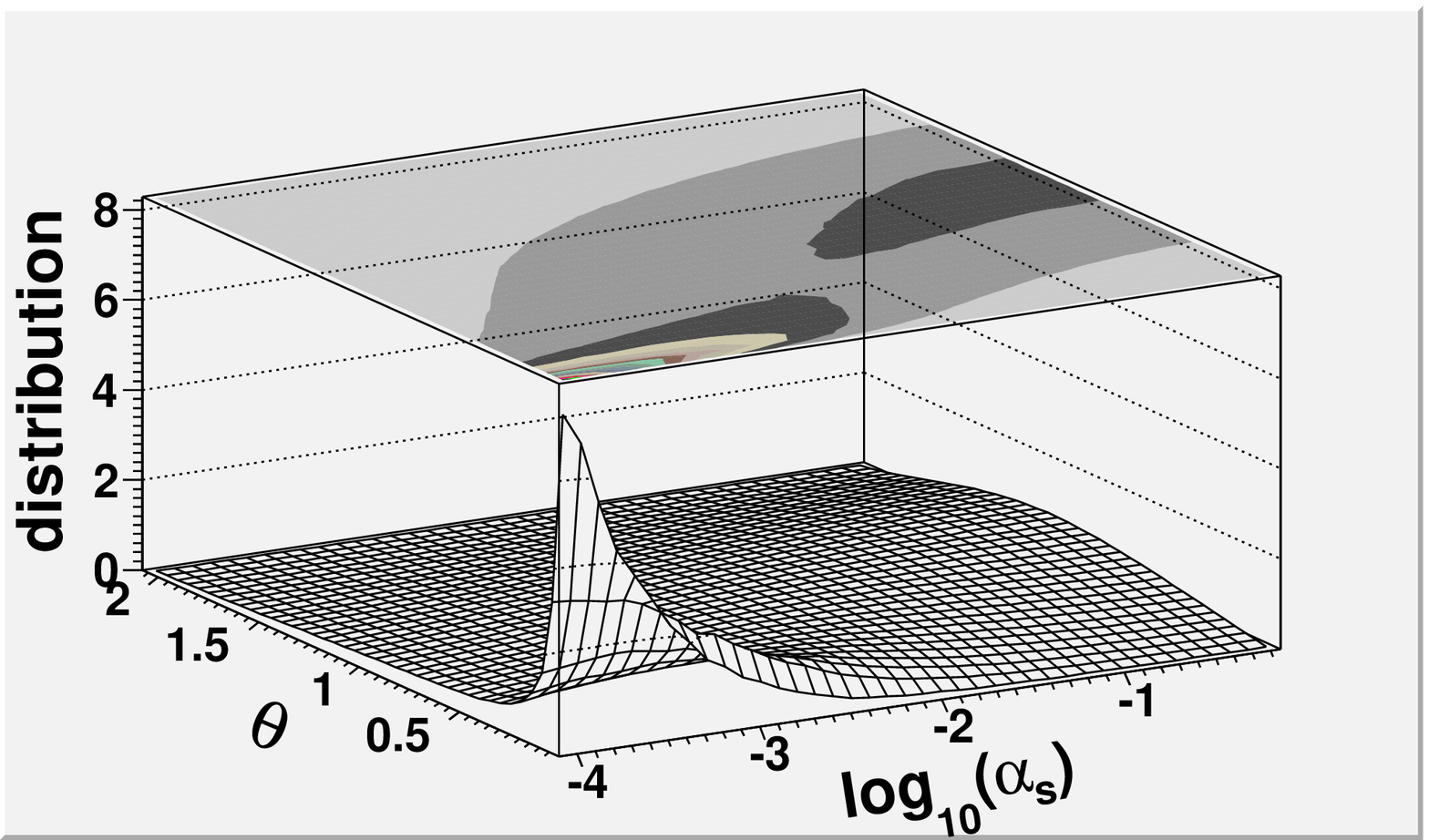} 
\includegraphics[width=0.3\textwidth]{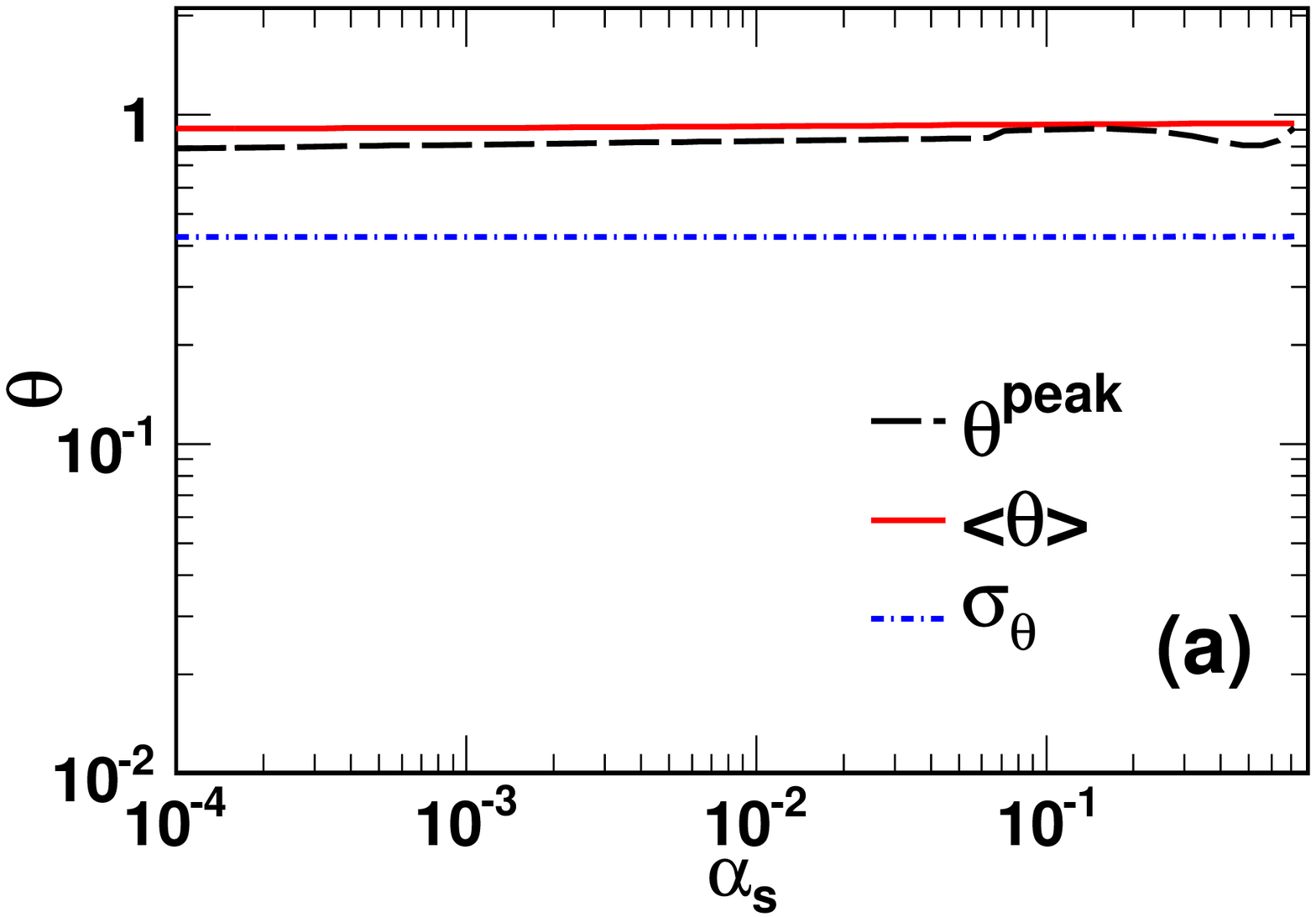} 
\includegraphics[width=0.3\textwidth]{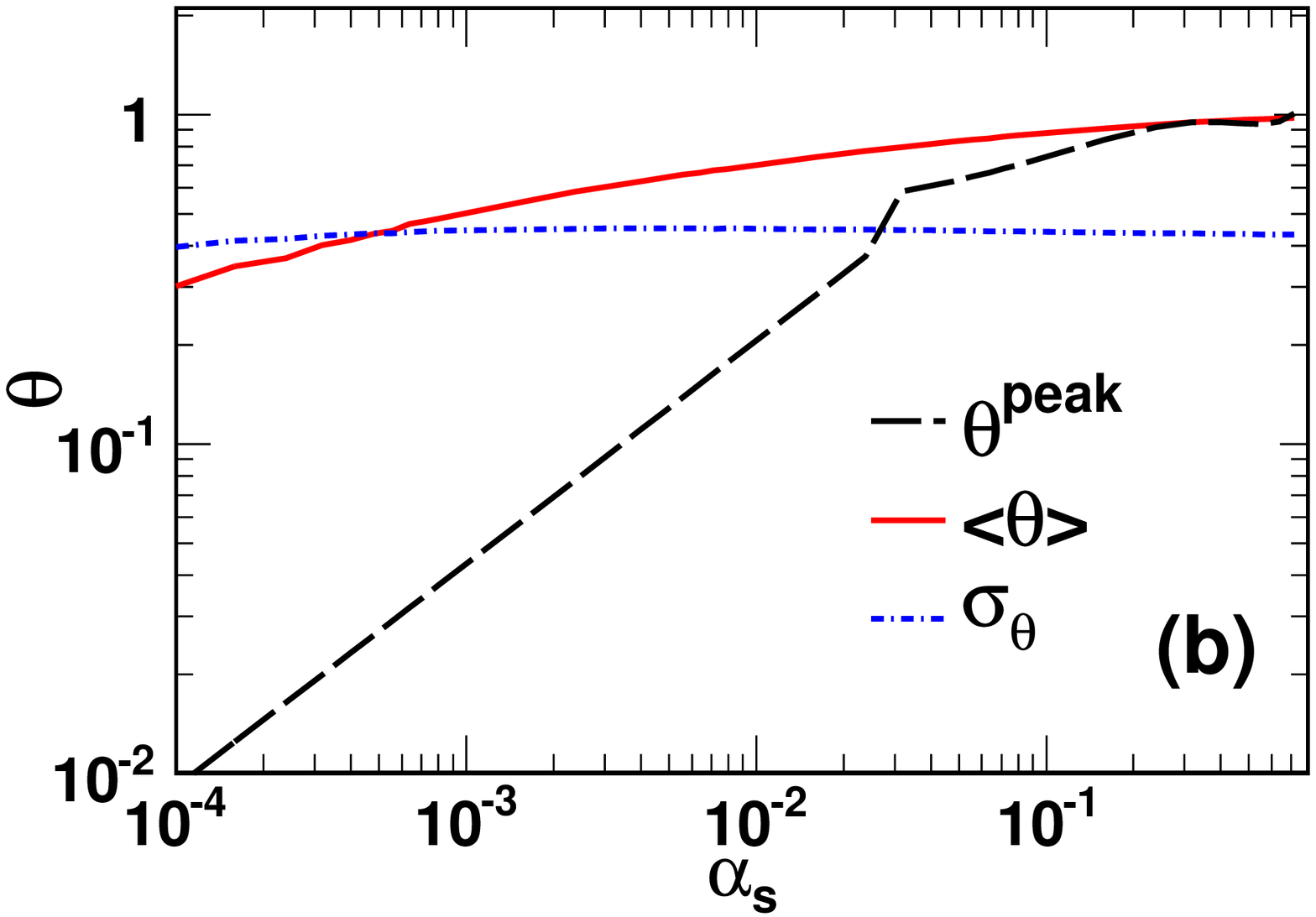} 
\includegraphics[width=0.3\textwidth]{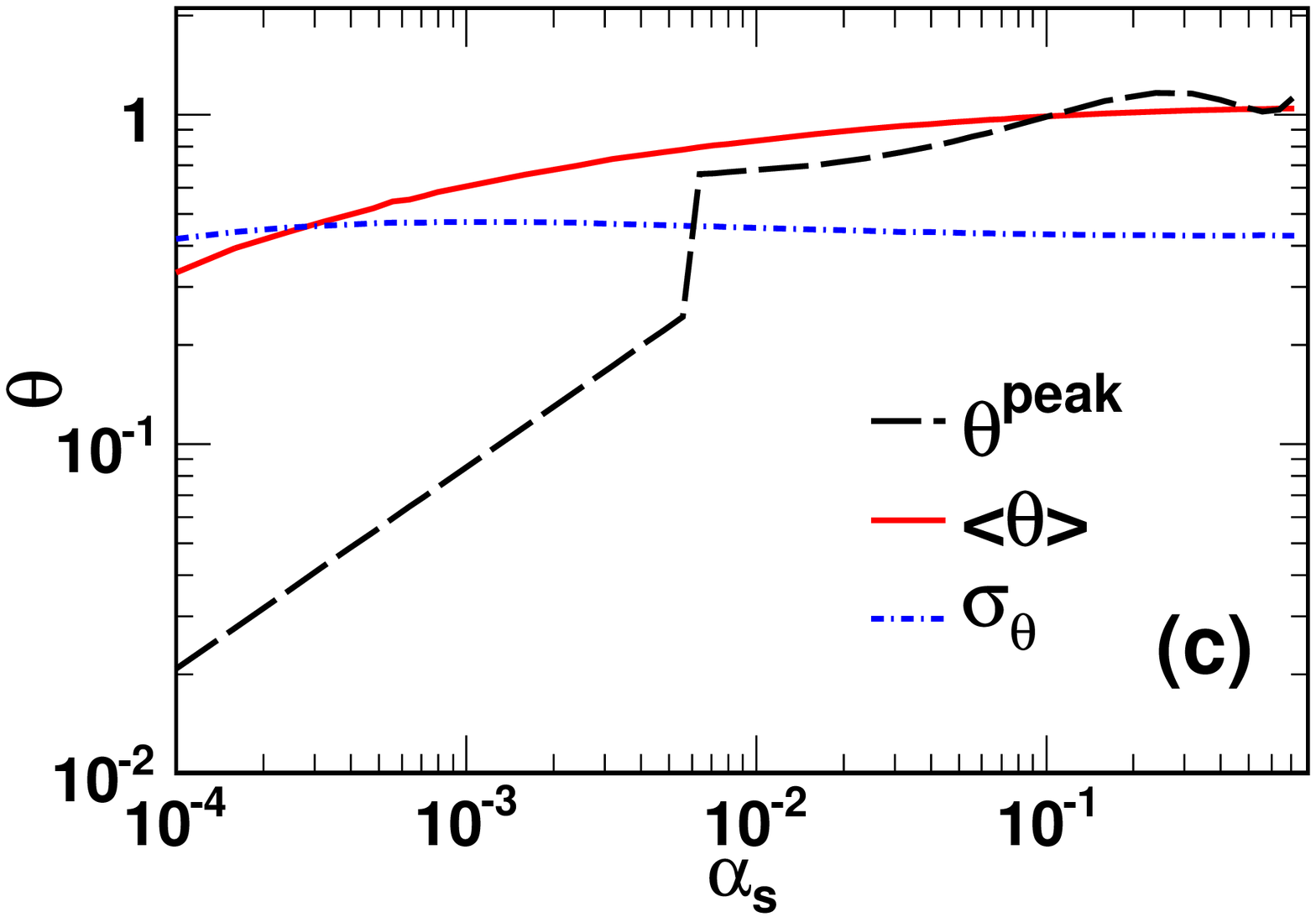} 
\end{figure}

\begin{figure}[tbp]
\caption{(Color online) Same as Fig. \protect\ref{fig:theta-lrf} but with $\protect\theta $
in the CM frame. }
\label{fig:theta-cm}
\includegraphics[width=0.3\textwidth]{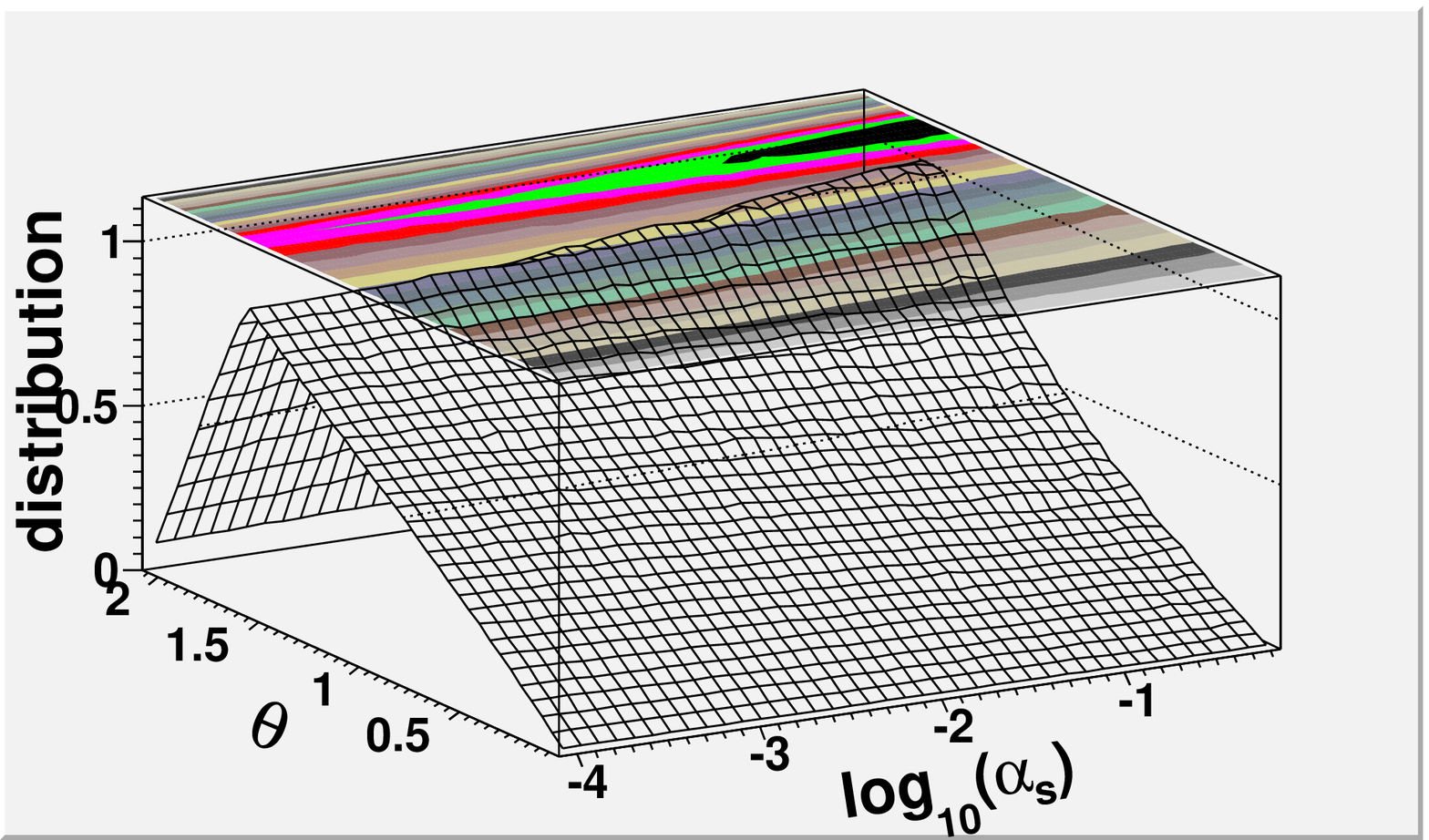} 
\includegraphics[width=0.3\textwidth]{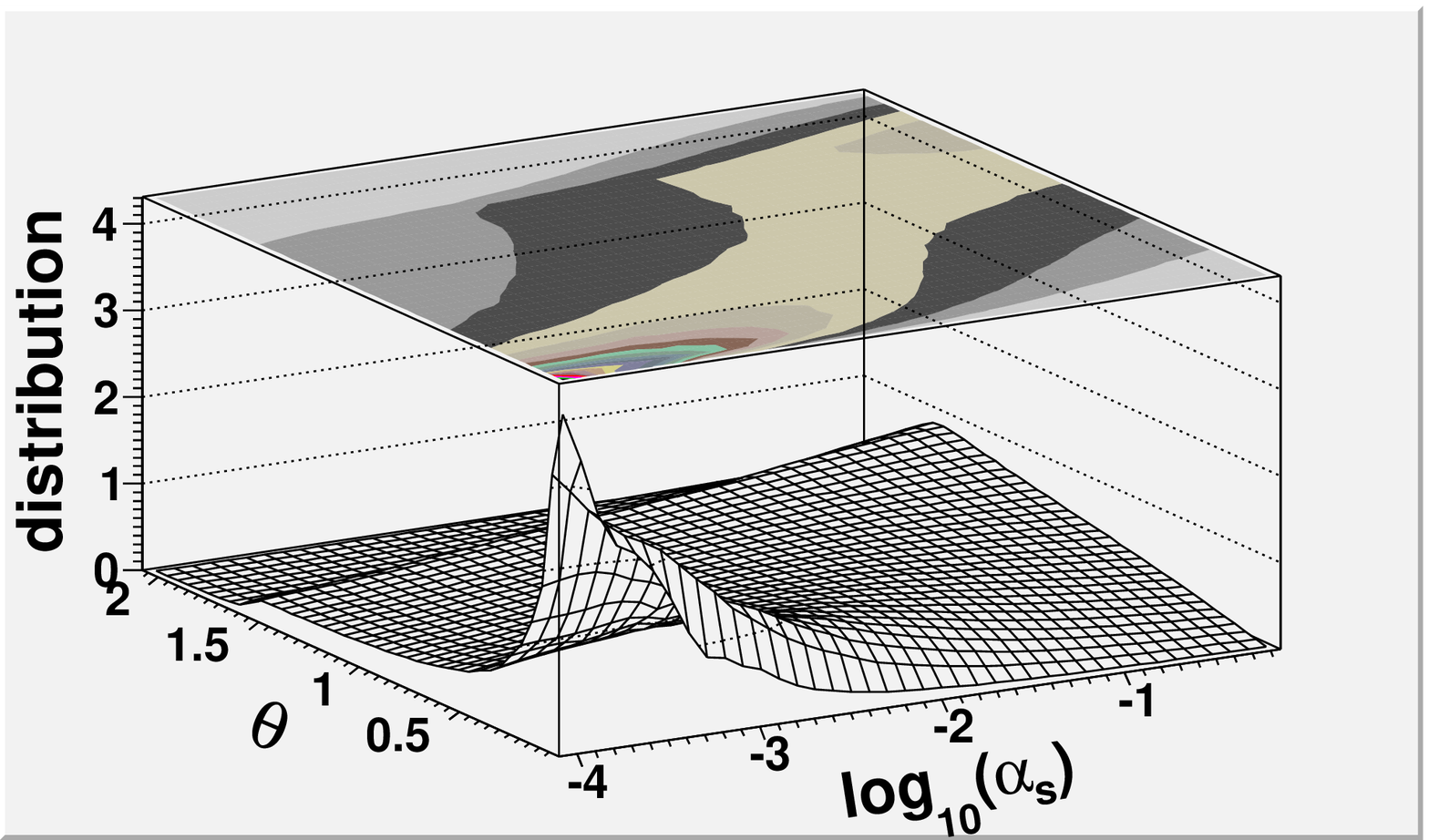} 
\includegraphics[width=0.3\textwidth]{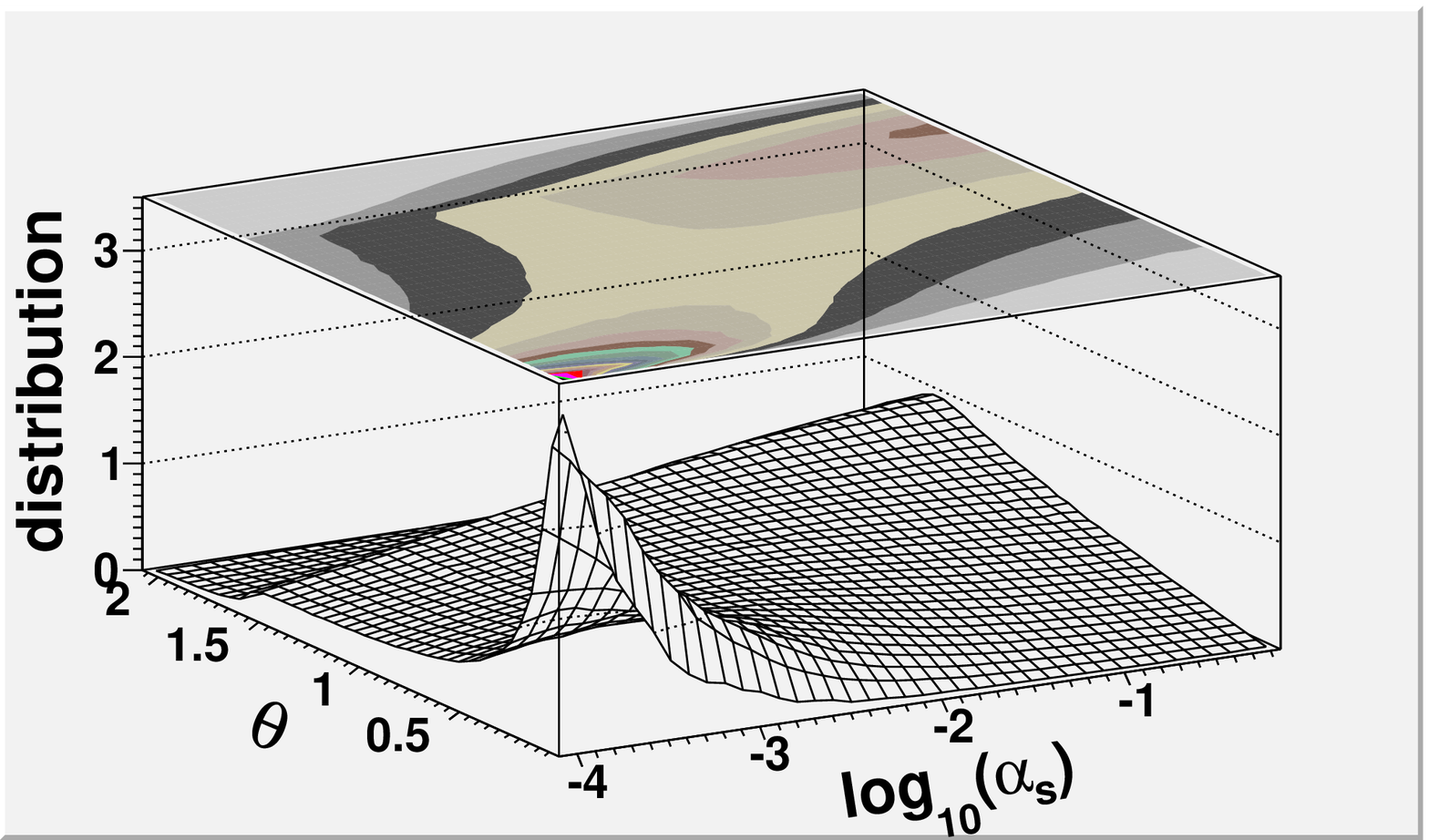} 
\includegraphics[width=0.3\textwidth]{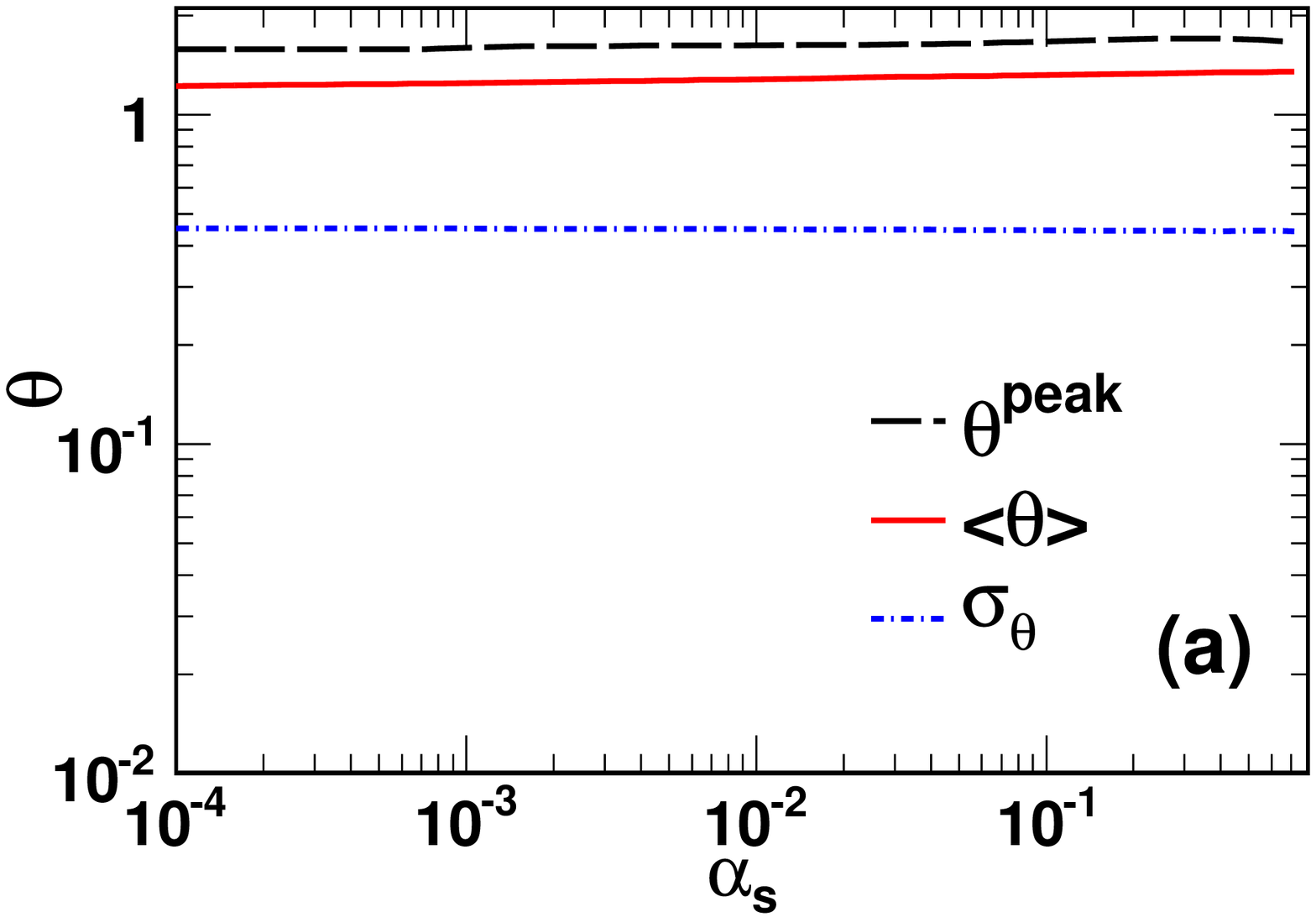} 
\includegraphics[width=0.3\textwidth]{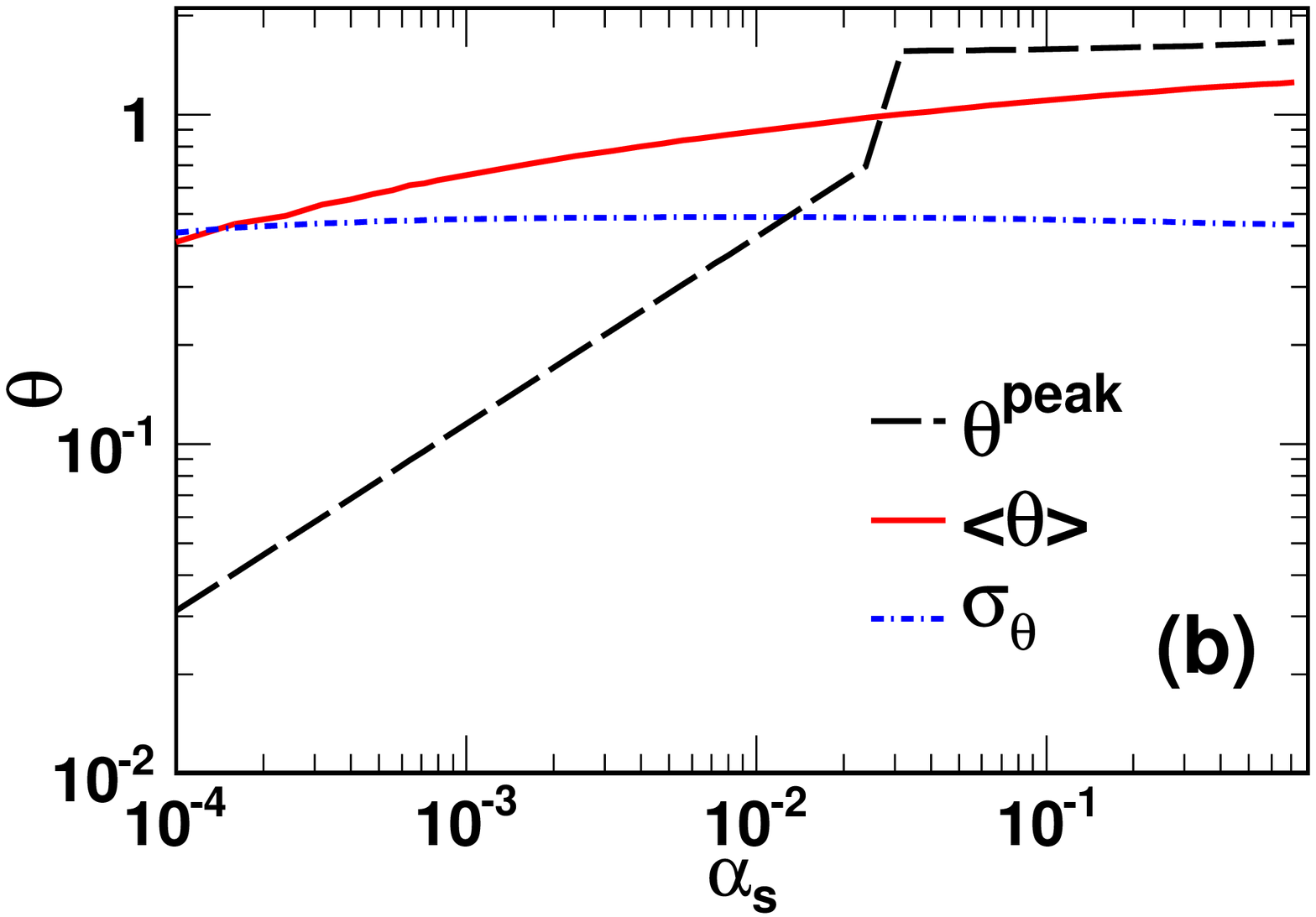}
\includegraphics[width=0.3\textwidth]{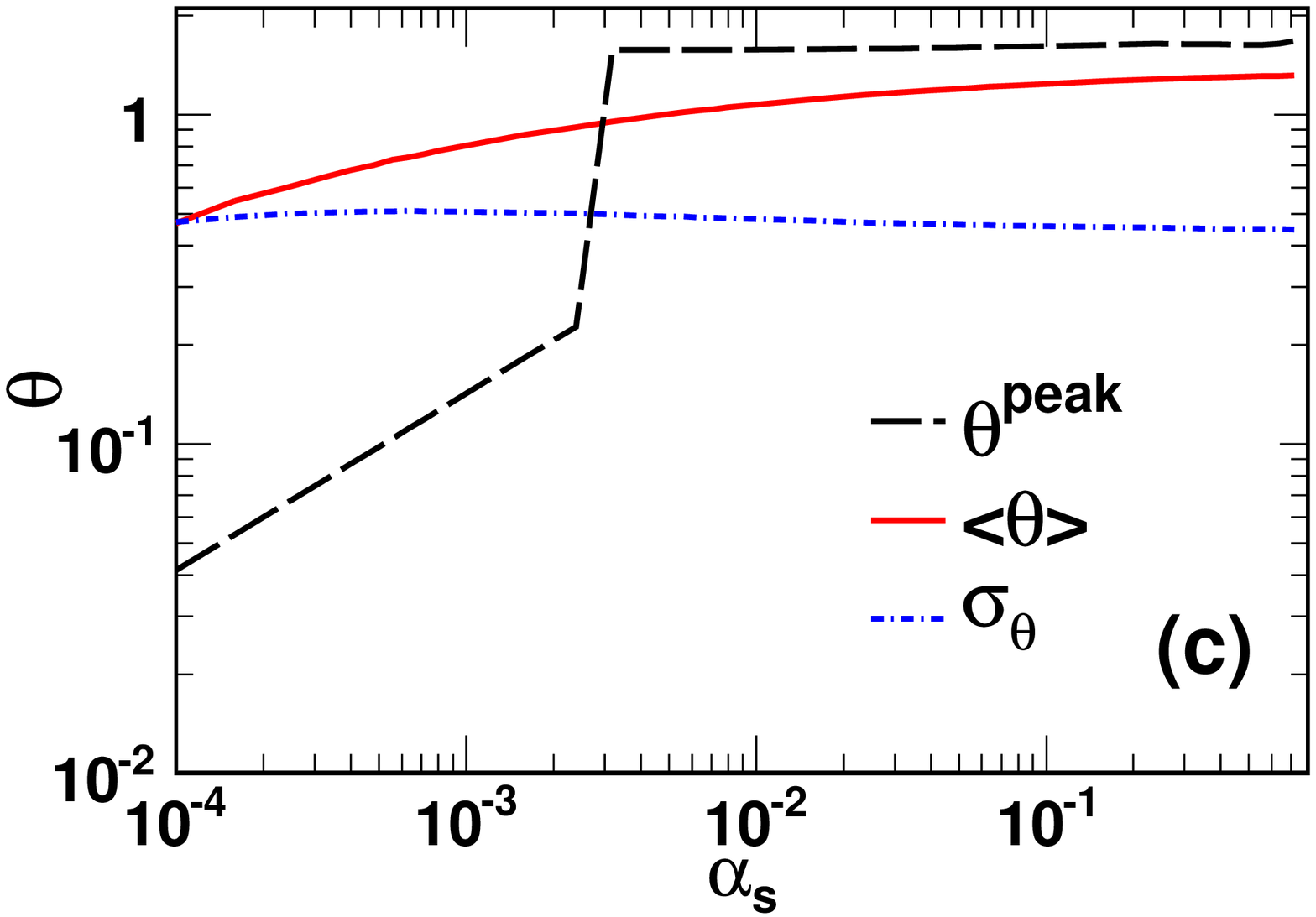}
\end{figure}

As mentioned in the introduction, although our result does not provide a
model independent check to AMY and ADM's results, we can still study the
angular correlation between final state gluons using our approach. Because
the 23 matrix element that we use is exact in vacuum, we can check, modulo
some model dependent medium effect, whether the correlation is dominated by
the near collinear splittings as asserted by AMY and ADM. 

We study the distribution of the minimum angle $\theta $ among the final
state gluons. If the near collinear splittings dominate, 
then the most probable configuration would be that two gluons whose angles are strongly
correlated and their relative angle tends to be the smallest among the three
relative angles in the final state. This can be seen most easily in the
center-of-mass (CM) frame of the 23 collision with two gluons going along
about the same direction while the third one is moving in the opposite
direction. We expect it is also the case in the fluid local rest frame. 

We find that  the distribution of $\theta $\ is peaked at $\theta ^{peak}\sim \sqrt{\alpha _{s}}$, 
analogous to the near collinear splitting asserted by AMY and ADM.
However, the average of $\theta $, $\langle \theta \rangle $, is
much bigger than its peak value, as its distribution is skewed with a long
tail. Below are more detailed descriptions of our results.

We show the distribution of $\theta $ in the fluid local rest frame in 
Fig.\ \ref{fig:theta-lrf}, and show the distribution in the CM frame of the 23
collision in Fig. \ref{fig:theta-cm}. In both figures, the left panel is the
distribution weighted by the phase space and the Bose-Einstein distribution
functions, the middle panel is weighted by the 23 contribution to $\eta $\
(denoted as $\eta _{23}$, which is the $\eta $\ analogy of the second term
in Eq. (\ref{eq:bulk3})) with the \textquotedblleft exact\textquotedblright\
matrix element, and the right panel is similar to the middle one with the GB
matrix element. 

We first look at the distribution in the fluid local rest frame in 
Fig. \ref{fig:theta-lrf}. The left panel plots do not depend on the interaction and
hence is $\alpha _{s}$\ independent. The distribution has $\theta
^{peak}\simeq \left\langle \theta \right\rangle $\ and the variation $\sigma
_{\theta }$\ is about the same size. In the middle panel, the $\eta _{23}$\
weighted distribution with the \textquotedblleft exact\textquotedblright\
matrix element, on the other hand, has $\theta ^{peak}\sim \sqrt{\alpha _{s}}$ 
at small $\alpha _{s}$, while $\left\langle \theta \right\rangle $ is
significantly bigger and $\sigma _{\theta }$ is close to its value in the
left panel. In the right panel, where the GB matrix element is used, $\theta ^{peak}$ 
is still close to be proportional to$\sqrt{\alpha _{s}}$ at small 
$\alpha _{s}$, but the angle is about twice as big as the \textquotedblleft
exact\textquotedblright\ case. 

The distribution in the 23 collision CM frame shown in 
Fig. \ref{fig:theta-cm} has a similar behavior as that in the fluid local rest frame
but the angles are in general much larger. 

The above analysis suggests that the GB formula, which takes the soft gluon
bremsstrahlung limit in the CM frame, still has some near collinear
splitting behavior in the fluid local rest frame. It is curious what the
nature of the long tail is. We will leave it for future investigation. 

\subsection{More aspects}

\begin{figure}[tbp]
\caption{(Color online)  Universal curves for $\zeta /s$ (this work), 
$\eta /s$ \cite{Chen:2010xk} and their ratio. These curves are universal
and suitable for a general $SU(N_{c})$\ pure gauge theory. }
\label{fig:Bulk2Shear}
\includegraphics[scale=0.4]{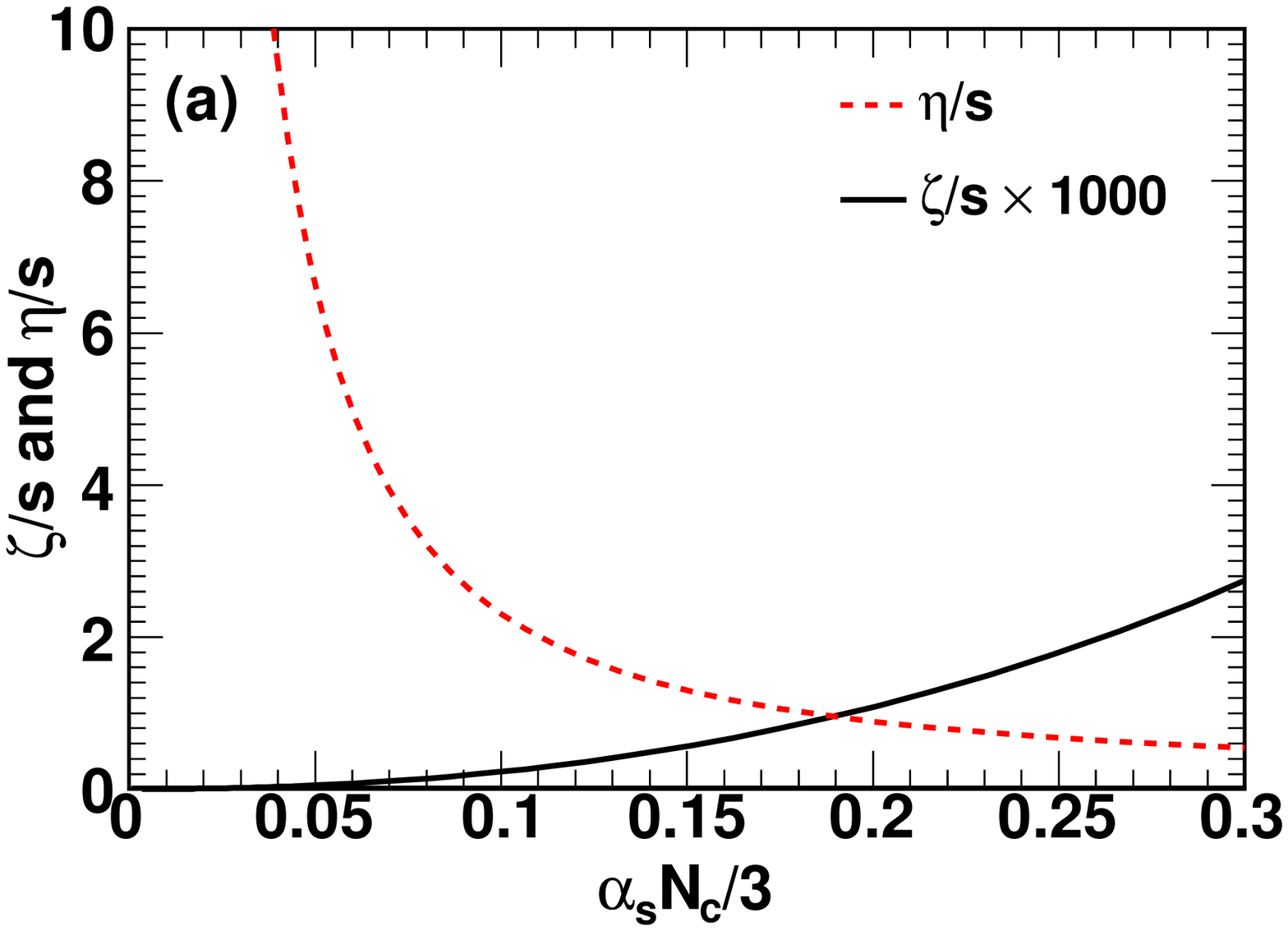} 
\includegraphics[scale=0.4]{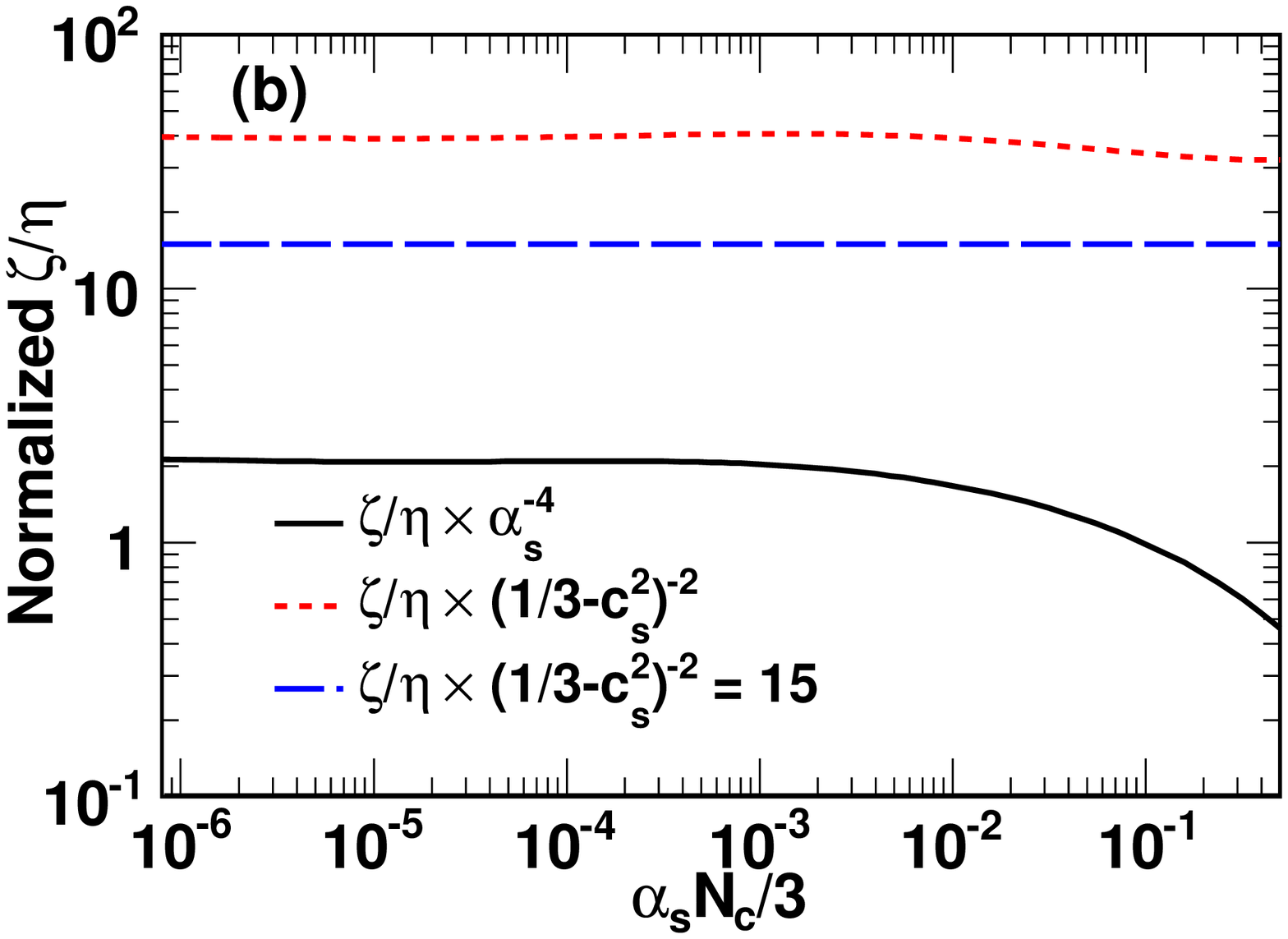}
\end{figure}

In Fig. \ref{fig:Bulk2Shear}, our results for $\zeta /s$ and $\eta /s$ ($%
\eta /s$ is computed in Ref. \cite{Chen:2010xk}) using the \textquotedblleft
exact\textquotedblright\ matrix element for the 23 process are shown in the
left panel and their ratio $\zeta /\eta $ in units of $\alpha _{s}^{4}$ and $%
\left( 1/3-c_{s}^{2}\right) ^{2}$ in the right panel. As we emphasize in
Sec. \ref{Nc}, these are universal curves suitable for a general $SU(N_{c})$
pure gauge theory.

The external gluon mass $m_{\infty }$ is included in the entropy density $s$
here, but it is a higher order effect and numerically very small at small $%
\alpha _s$. In the range where perturbation theory is reliable ($\alpha
_{s}\lesssim 0.1$), $\zeta $ is always smaller than $\eta $ by at least
three orders of magnitude. One can see that our result of $\zeta /\eta $
agrees with $15\left( 1/3-c_{s}^{2}\right) ^{2}$ of Weinberg parametrically 
\cite{Weinberg:1971mx}, and it is rather close to the LL one in Eq. (\ref%
{LL-zeta-eta}).

\begin{figure}[tbp]
\caption{(Color online) (a) $\eta /s$ for a pion gas 
\cite{Chen:2006iga} and a gluon plasma with LQCD 
\cite{Meyer:2007ic,Meyer:2009jp,Nakamura:2004sy} and perturbative QCD 
\cite{Chen:2010xk}, together with $\eta /s$ extracted from
RHIC elliptical flow ($v_{2}$): I \cite{Luzum:2008cw} and II 
\cite{Song:2008hj}. The arrow below the line of "Hydro+$v_{2}$ data
II" indicates that it is an upper bound. (b) $\zeta /s$ for
a massive \cite{Lu:2011df,FernandezFraile:2008vu} and massless \cite{Chen:2006iga}
pion gas and a gluon plasma with LQCD \cite{Meyer:2010ii} and perturbative QCD (this
work). The sum rule result \cite{Karsch:2007jc} is for $N_{f}=3$. 
The massive pion curves are denoted as "pion gas I" 
(inelastic process \cite{Lu:2011df}) and 
"pion gas II" (elastic process \cite{FernandezFraile:2008vu}). }
\label{fig:Tc}
\includegraphics[scale=0.4]{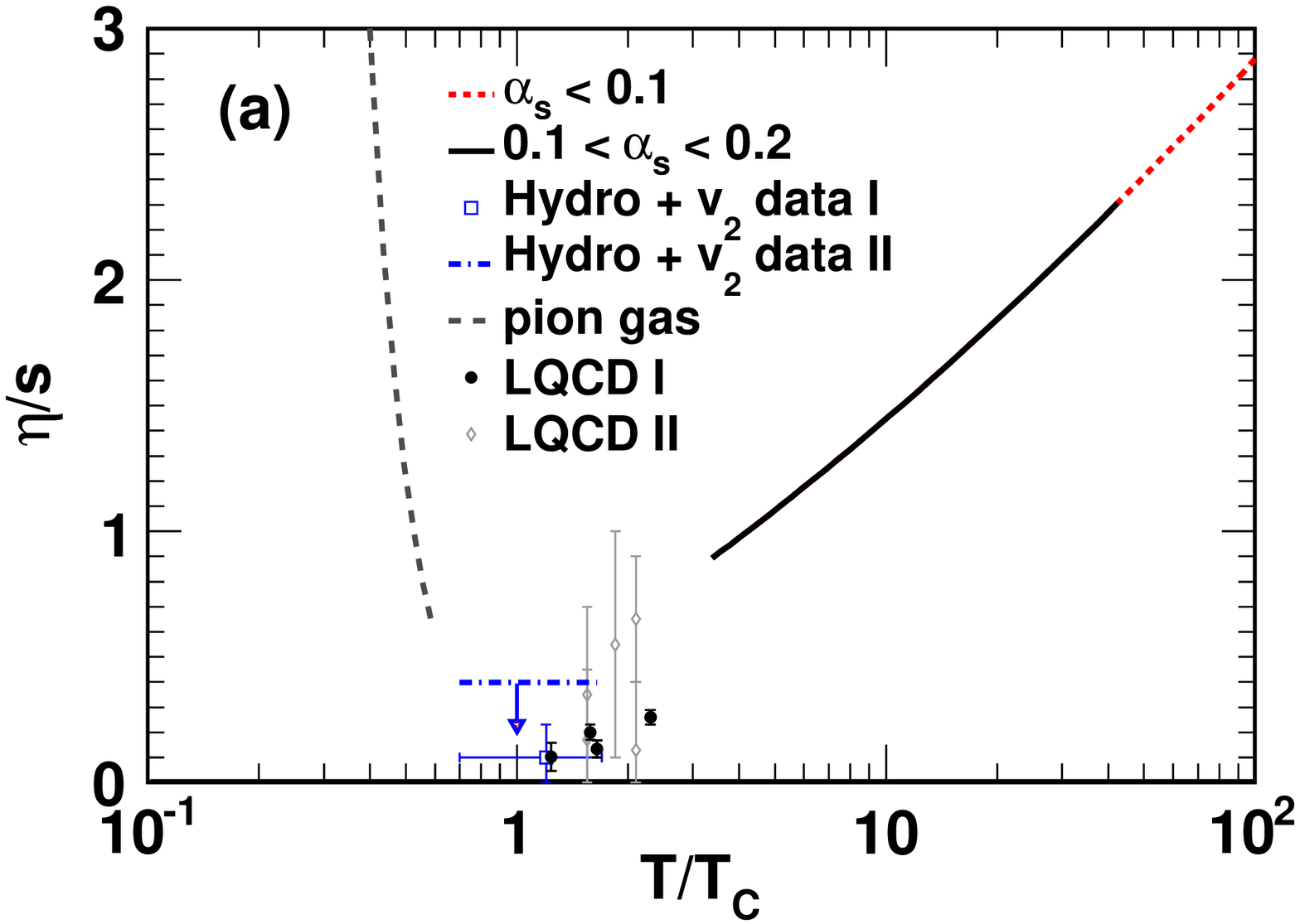} 
\includegraphics[scale=0.4]{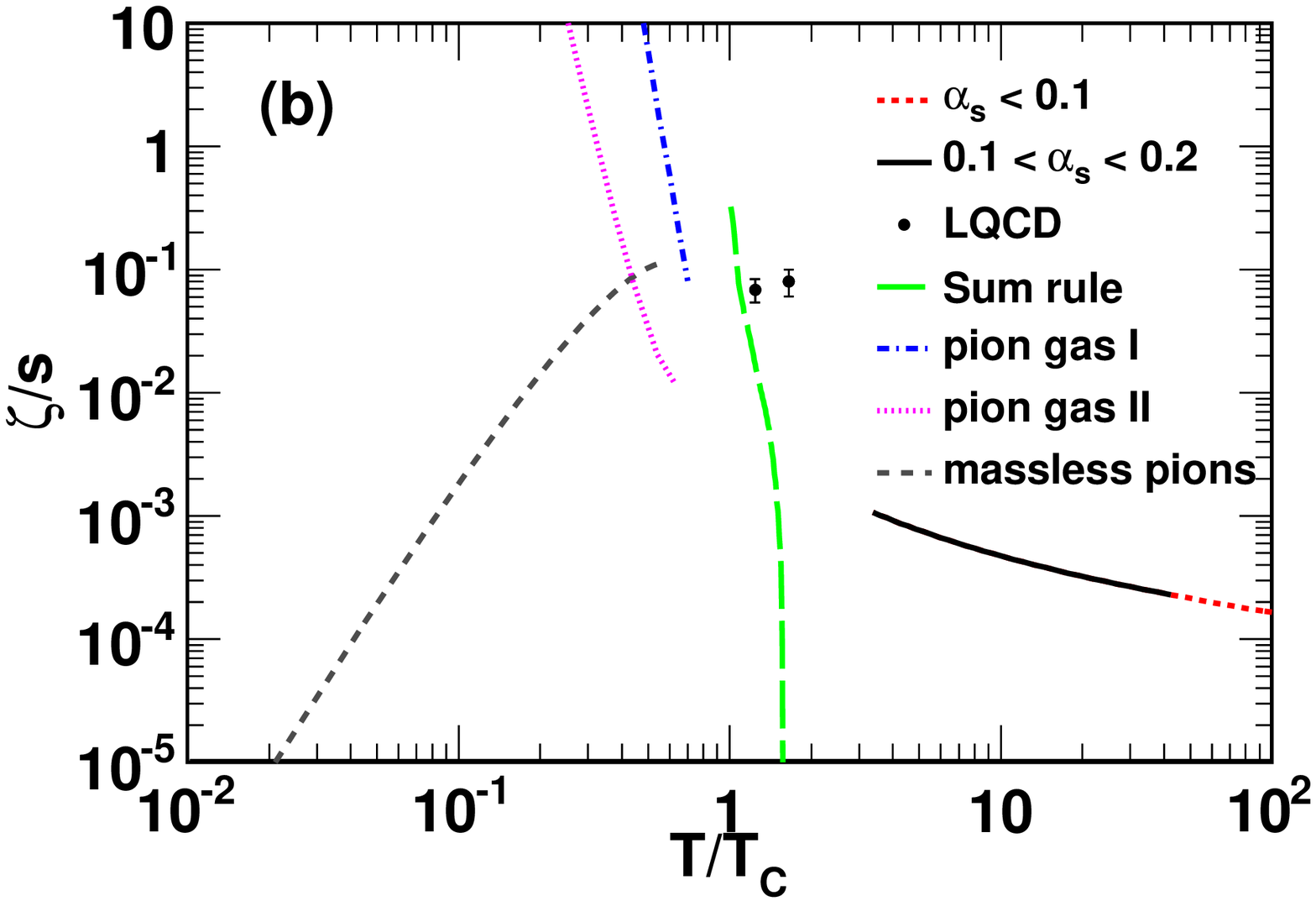}
\end{figure}

In Fig. \ref{fig:Tc}, we have plotted $\eta /s$ vs. $T/T_{c}$ and $\zeta /s$
vs. $T/T_{c}$ for QCD with various number of light quark flavors $N_{f}$
(and different $T_{c}$'s are used in different systems) at zero baryon
chemical potential. In $\eta /s$, $T/T_{c}\ll 1$, the QCD result is
calculated by the pion gas system using the Boltzmann equation \cite%
{Chen:2006iga} (the kaon mass is more than two time bigger than $T_{c}$
---too heavy to be important for $T/T_{c}\ll 1$; for other calculations in
hadronic gases, see \cite%
{Prakash:1993bt,Dobado:2003wr,Dobado:2001jf,Chen:2007xe,Itakura:2007mx}).
The $T/T_{c}\gtrsim 1$ result is for gluon plasma using lattice QCD (LQCD) 
\cite{Meyer:2007ic,Meyer:2009jp,Nakamura:2004sy} (see \cite{Meyer:2011gj}
for a recent review; for a lattice inspired model around $T_{c}$, see e.g.
Ref. \cite{Hidaka:2009ma}). This result has assumed a certain functional
form for the spectral function and hence has some model dependence. Note
that in this temperature region, there might be anomalous shear viscosity
arising from coherent color fields in the early stage of the QGP \cite%
{Asakawa:2006tc}. We have also shown the value of $\eta /s$ extracted from
the elliptic flow ($v_{2}$) data of RHIC using hydrodynamics: $\eta
/s=0.1\pm 0.1(\mathrm{theory})\pm 0.08(\mathrm{experiment})$ \cite%
{Luzum:2008cw} (denoted as \textquotedblleft Hydro$+v_{2}$ data
I\textquotedblright ) and $\eta /s<5\times 1/(4\pi )$ \cite{Song:2008hj}
(denoted as \textquotedblleft Hydro$+v_{2}$ data II\textquotedblright ). And
we have assigned a conservative temperature range $T=0.24\pm 0.10$ GeV that
covers the initial and final temperatures in the hydrodynamic evolution ($%
T_{f}=0.14$\ GeV, $T_{i}\lesssim 0.34$\ GeV).

For $T/T_{c}\gg 1$, we use the perturbative result of the gluon plasma with
the 22 and 23 processes in the Boltzmann equation \cite{Chen:2010xk} and the
standard two-loop renormalization (the scheme dependence is of higher order)
for the SU(3) pure gauge theory 
\begin{equation}
\frac{1}{4\pi \alpha _{s}\left( T\right) }=2\beta _{0}\ln \left( \frac{\mu T%
}{\Lambda _{\overline{MS}}}\right) +\frac{\beta _{1}}{\beta _{0}} \ln \left(
2\ln \left( \frac{\mu T}{\Lambda _{\overline{MS}}}\right) \right) ,
\end{equation}
where $\beta _{0}=11/(16\pi ^{2})$ and $\beta _{1}=102/(16\pi ^{2})^{2}$.
Fitting to lattice data at $1.2\lesssim T/T_{c}\lesssim 2$ yields $\mu
\simeq 1.14\pi $, $\Lambda _{\overline{MS}}\simeq 261$ MeV and $T_{c}\simeq
202$ MeV \cite{Kaczmarek:2005ui}. When $T/T_{c}\simeq 3.3$ and $42$, $\alpha
_{s}=0.2$ and $0.1$, respectively. If $\eta /s$ above $T_{c}$ is dominated
by the gluon contribution so the gluon plasma result ($N_{f}=0$) is close to
that of $N_{f}=3$ QCD \footnote{It is curious how to compute $\eta /s$ below $T_{c}$ for 
$N_{f}=0$ and $1$. There is no Goldstone mode in this case and there is no
obvious gap in the spectrum to justify an effective field theory treatment.
The lattice QCD computation also suffers from small correlator signals due
to heavy hadron masses in the intermediate states.} , then Fig. \ref{fig:Tc} shows that $\eta /s$ might
have a local minimum at $T_{c}$ \cite{Csernai:2006zz,Chen:2006iga,Bluhm:2010qf} .

For $\zeta /s$ with $T/T_{c}\ll 1$, the QCD result is calculated by the Boltzmann
equation for massless \cite{Chen:2007kx}\ and massive 
\cite{Lu:2011df,FernandezFraile:2008vu} 
(also in Ref. \cite{Dobado:2011qu,Chakraborty:2010fr}) pions. 
For massless pions, $\zeta /s$ is increasing in $T$ since 
it is expected when the pion self-coupling vanishes
(or equivalently the pion decay constant $f_{\pi }\rightarrow \infty $), $%
\zeta $ also vanishes. Thus, the dimensionless combination $\zeta /s\propto
\left( T/f_{\pi }\right) ^{z}$, where $z$ is some positive number. 
For massive pions, the expected non-relativistic limit for the bulk
viscosity reads \cite{FernandezFraile:2008vu} $\zeta \sim f^4_{\pi}\sqrt{T} / m_\pi^{3/2}$, 
where $m_\pi =$138 MeV is the physical pion mass and one uses 
Weinberg's low-energy result for the pion-pion cross section at low
energy (low temperature) [72]. This suggests the (non-relativistic)
conformal symmetry is recovered at zero $T$ when particle number
conservation is imposed. In the relativistic case, Ref.\ \cite{Lu:2011df} argues
that the number changing process (the 24 process, 23 not allowed by
parity conservation) is slower than 22, so it controls the time scale
for the system to go back to thermal equilibrium. 
At low enough $T$, this time scale
is very long since there are not many pions energetic enough to collide and
produce four pions. However, if the time scale is longer than that of the fire ball
expansion at RHIC, the elastic scattering \cite{FernandezFraile:2008vu,Prakash:1993bt,Davesne:1995ms} 
(see also \cite{NoronhaHostler:2008ju,Paech:2006st}) 
is more relevant phenomenologically. 
For $T/T_{c}\gtrsim 1$, lattice QCD calculation of a gluon plasma is shown 
\cite{Meyer:2010ii} together with the sum rule result with 
$N_{f}=3$ \cite{Kharzeev:2007wb,Karsch:2007jc}. Both of them have some model dependence on
the shape of the spectral function used. This issue was discussed
extensively in Refs. \cite{Teaney:2006nc,Moore:2008ws,Romatschke:2009ng}
which inspired Ref. \cite{Meyer:2010ii} to include a delta function
contribution to the spectral function which was missed in the earlier result
of Ref. \cite{Meyer:2007dy}. The same delta function will modify the sum
rule result \cite{Kharzeev:2007wb,Karsch:2007jc} as well. This is yet to be
worked out.

For $T/T_{c}\gg 1$, the perturbative result of the gluon plasma calculated
in this work is shown. We see that although $\zeta /s$ for the massive pion
case is decreasing in $T$ for small $T$. It should merge to the massless
pion result when the pion thermal energy $\sim 3T$ is bigger than $m_{\pi }$%
. Thus, it is still possible that $\zeta /s$ has a local maximum at $T_{c}$
as in some model calculations \cite{Gubser:2008yx,Li:2009by,Bluhm:2010qf}
provided there is no much difference between the $N_{f}=0$ and $N_{f}=3$
results above $T_{c}$.

It is very interesting that $\zeta >\eta $ for the gluon plasma just above $%
T_{c}$. This suggests a fluid could still be perfect without being
conformal, like the AdS/CFT model of Ref. \cite{Gubser:2008yx}. Finally, it
is intriguing that $\eta /s$ might have a local minimum at $T_{c}$ and 
$\zeta /s$ might have a local maximum at $T_{c}$. However, despite there are
many other systems exhibiting this behavior for $\eta /s$ \cite%
{Csernai:2006zz,Lacey:2006bc,Chen:2007xe,Chen:2007jq}, there are
counterexamples showing that it is not universal \cite%
{Chen:2010vf,Chen:2010vg,Dobado:2009ek,FernandezFraile:2010gu}.

\section{Conclusions}

We have calculated the shear and bulk viscosity of a weakly interacting
gluon plasma with 22 and 23 collisional processes and a simple treatment to model the
LPM effect. Our results agree with the results of AMY and ADM within errors. By
studying the 23 contribution to $\eta $, we find that the minimum angle $\theta $ 
among the final state gluons has a distribution that is peaked at 
$\theta \sim \sqrt{\alpha _{s}}$, analogous to the near collinear splitting
asserted by AMY and ADM. However, the average of $\theta $ is much bigger
than its peak value, as its distribution is skewed with a long tail which is
worth further exploration. The same $\theta $ behavior is also seen if the 23
matrix element is taken to the soft gluon bremsstrahlung limit in the CM
frame. This suggests that the soft gluon bremsstrahlung in the CM frame still has
some near collinear behavior in the fluid local rest frame. We
also generalize our result to a general $SU(N_{c})$ pure gauge theory and
summarize the current theoretical results for viscosities in QCD.

\vspace{0.3cm}

Acknowledgement: QW thanks C. Greiner for bringing our attention to their
latest results about the shear viscosity for the 23 process. 
We thank G. Moore for useful communications related to his work. 
JWC thanks INT, Seattle, for hospitality.  
JWC is supported by the NSC, NCTS, and CASTS of ROC. 
QW is supported in part by the National Natural Science
Foundation of China (NSFC) under grant 10735040 and 11125524. 
HD is supported in part by the NSFC under grant 11105084 and the
Natural Science Foundation of the Shandong province under grant ZR2010AQ008.
JD is supported in part by the NSFC under grant 11105082 and 
the Innovation Foundation of Shandong University
under grant 2010GN031. 

\appendix

\section{Soft gluon bremsstrahlung}

\label{app1}

In this appendix, we give the details of the derivation of the GB formula or
the matrix element for the soft gluon bremsstrahlung. We work in the CM
frame of the initial or final state where the longitudinal direction is
defined as that of $\mathbf{p}_{1}$ or $\mathbf{p}_{2}$. The conditions for
the soft gluon bremsstrahlung are: $s\gg p_{iT}$ and $k_{T}\gg yq_{T}$ (or $%
s\rightarrow \infty $ and $y\rightarrow 0$). This means the energy of the
bremsstrahlung gluon, say $E_{5}$, is much smaller than other two gluons, $%
E_{5}\ll E_{3},E_{4}$.

It is convenient to use the Mandelstam-like variables defined as 
\begin{eqnarray}
&&s=(p_{1}+p_{2})^{2},\;t=(p_{1}-p_{3})^{2},\;u=(p_{1}-p_{4})^{2},  \notag \\
&&s^{\prime }=(p_{3}+p_{4})^{2},\;t^{\prime }=(p_{2}-p_{4})^{2},\;u^{\prime
}=(p_{2}-p_{3})^{2},  \notag \\
&&T_{i5}=(k_{i}+k_{5})^{2},\;\;(i=1,2,3,4).
\end{eqnarray}%
Here we assume all gluons are massless, so we obtain 
\begin{eqnarray}
&&(12)=\frac{s}{2},\;(13)=-\frac{t}{2},\;(14)=-\frac{u}{2},  \notag \\
&&(23)=-\frac{u^{\prime }}{2},\;(24)=-\frac{t^{\prime }}{2},\;(34)=\frac{%
s^{\prime }}{2},  \notag \\
&&(15)=\frac{T_{15}}{2},\;(25)=\frac{T_{25}}{2},\;(35)=\frac{T_{35}}{2}%
,\;(45)=\frac{T_{45}}{2}.
\end{eqnarray}%
Using light-cone variables in Eq. (\ref{light-cone}) and taking the limit $%
s\rightarrow \infty $ or $s\gg p_{iT}^{2}$, we have 
\begin{eqnarray}
t &=&-\frac{1}{1-y}(\mathbf{q}_{T}-\mathbf{k}_{T})^{2},  \notag \\
u &\approx &-s,  \notag \\
s^{\prime } &\approx &(1-y)s,  \notag \\
t^{\prime } &=&-q_{T}^{2},  \notag \\
u^{\prime } &=&-(1-y)s,  \notag \\
T_{15} &\approx &k_{T}^{2}/y,  \notag \\
T_{25} &\approx &ys,  \notag \\
T_{35} &\approx &\frac{(\mathbf{k}_{T}-y\mathbf{q}_{T})^{2}}{(1-y)y},  \notag
\\
T_{45} &\approx &ys.
\end{eqnarray}%
We see that $t,t^{\prime },T_{i5}(i=1,2,3,4)$ are small. In evaluating $%
\left\vert M_{12\rightarrow 345}\right\vert ^{2}$, we denote $(12345)\equiv
1/[\left( 12\right) \left( 23\right) \left( 34\right) \left( 45\right)
\left( 51\right) ]$, and we can evaluate all quantities in the denominator
of Eq. (\ref{eq:m3}), 
\begin{eqnarray}
(12345) &=&-\frac{1}{su^{\prime }s^{\prime }T_{15}T_{45}},  \notag \\
(12354) &=&\frac{1}{su^{\prime }uT_{35}T_{45}},  \notag \\
(12435) &=&-\frac{2^{5}}{ss^{\prime }t^{\prime }T_{15}T_{35}},  \notag \\
(12453) &=&\frac{1}{stt^{\prime }T_{35}T_{45}},  \notag \\
(12534) &=&-\frac{1}{ss^{\prime }uT_{25}T_{35}},  \notag \\
(12543) &=&-\frac{1}{ss^{\prime }tT_{25}T_{45}},  \notag \\
(13245) &=&-\frac{1}{u^{\prime }tt^{\prime }T_{15}T_{45}},  \notag \\
(13254) &=&-\frac{1}{uu^{\prime }tT_{25}T_{45}},  \notag \\
(13425) &=&\frac{1}{s^{\prime }tt^{\prime }T_{15}T_{25}},  \notag \\
(13524) &=&-\frac{1}{utt^{\prime }T_{25}T_{35}},  \notag \\
(14235) &=&-\frac{1}{uu^{\prime }t^{\prime }T_{15}T_{35}},  \notag \\
(14325) &=&\frac{1}{uu^{\prime }s^{\prime }T_{15}T_{25}},
\end{eqnarray}%
where we have factored out $2^{5}$. Note that other permutations which do
not appear are given by the identity $(12345)=(15432)$. Then we can collect
the most singular parts involving $t,t^{\prime },T_{i5}(i=1,2,3,4)$ in the
denominator and obtain 
\begin{eqnarray}
\left\vert M_{12\rightarrow 345}\right\vert ^{2} &\sim &\frac{1}{stt^{\prime
}T_{35}T_{45}}-\frac{1}{u^{\prime }tt^{\prime }T_{15}T_{45}}+\frac{1}{%
s^{\prime }tt^{\prime }T_{15}T_{25}}-\frac{1}{utt^{\prime }T_{25}T_{35}}, 
\notag \\
&&-\frac{1}{ss^{\prime }t^{\prime }T_{15}T_{35}}-\frac{1}{ss^{\prime
}tT_{25}T_{45}}-\frac{1}{uu^{\prime }tT_{25}T_{45}}-\frac{1}{uu^{\prime
}t^{\prime }T_{15}T_{35}},  \notag \\
&\sim &\frac{2}{s^{2}q_{T}^{2}}\left[ \frac{1}{(\mathbf{q}_{T}-\mathbf{k}%
_{T})^{2}}\frac{(1-y)^{2}}{(\mathbf{k}_{T}-y\mathbf{q}_{T})^{2}}+\frac{1}{(%
\mathbf{q}_{T}-\mathbf{k}_{T})^{2}}\frac{1}{k_{T}^{2}}+\frac{y^{2}}{(\mathbf{%
k}_{T}-y\mathbf{q}_{T})^{2}k_{T}^{2}}\right] .
\end{eqnarray}%
One can see that the matrix element squared has singularities from three
poles at $k_{T}^{2}=0$, $(\mathbf{q}_{T}-\mathbf{k}_{T})^{2}=0$. For the
soft limit, $k_{T}\gg yq_{T}$, this can be realized by setting $y\rightarrow
0$, we obtain 
\begin{equation}
\left\vert M_{12\rightarrow 345}\right\vert _{soft}^{2}\sim \frac{4}{%
s^{2}q_{T}^{2}}\frac{1}{(\mathbf{q}_{T}-\mathbf{k}_{T})^{2}k_{T}^{2}},
\end{equation}%
which reproduces the GB formula.

\section{Error Estimation}

\label{app2}

The error bands of $\eta $ and $\zeta $ shown in Figs. \ref{fig:shear23} and %
\ref{fig:bulk23} are based on the estimation of the following errors:

(a) HTL corrections for the 23 process: In the 22 process, if we replace the
HTL scattering amplitude of Eq. (\ref{eq:htl}) by that of Eq. (\ref{MD})
with $m_{D}$ as the regulator, then the 22 collision rate is reduced by $%
\sim 30\%$ for $\alpha _{s}\simeq 0.005$-$0.1$. At smaller $\alpha _{s}$,
the effect becomes smaller and eventually becomes negligible at $\alpha
_{s}=10^{-8}$. The reduction arises because the HTL magnetic screening
effect gives a smaller IR cut-off than $m_{D}$. Analogously, using $m_{D}$
as the regulator in the 23 process tends to under-estimate the 23 collision
rate and gives a larger $\eta $ and $\zeta $.

(b) LPM effect: Our previous calculation on $\eta $ using the Gunion-Bertsch
formula shows that implementing the $m_{D}$\ regulator gives a very close
result to the LPM effect \cite{Chen:2009sm}. Thus, we will estimate the size
of the LPM effect by increasing the external gluon mass $m_{g}$ from $%
m_{\infty }$ to $m_{D}$.

(c) Higher order effect: The higher order effect is parametrically
suppressed by $O(\sqrt{\alpha _{s}})$, but the size is unknown. Computing
this effect requires a treatment beyond the Boltzmann equation \cite%
{Hidaka:2010gh} and the inclusion of the 33 and 24 processes. We just
estimate the effect to be $\sqrt{\alpha _{s}}$ times the leading order which
is $\sim 10\%$ at $\alpha _{s}=0.01$. (Note that we estimated the higher
order effect to be $O(\alpha _{s})$\ suppressed in Ref. \cite{Chen:2009sm}.
But since the expansion parameter in finite temperature field theory is $g$\
instead of $g^{2}$, we enlarge the error here.)

Combining the above analyses, we consider errors from (a) to (c). 
To compute a recommended range of $\zeta $ (the range of $\eta $\ is
computed analogously), we will work with the $R_{22}$ and $R_{23}$ collision
rates defined as 
\begin{eqnarray}
R_{23}^{-1} &\equiv &\zeta _{23},  \notag \\
\left( R_{22}+R_{23}\right) ^{-1} &\equiv &\zeta _{22+23},
\end{eqnarray}%
where $\zeta _{23}$ is the bulk viscosity for a collision with the 23
process only. Using HTL instead of $m_{D}$ for the gluon propagator enhances
the 22 rate by a factor of 
\begin{equation}
\delta \equiv \frac{R_{22(HTL)}}{R_{22(MD)}}.
\end{equation}%
We will assume that the same enhancement factor appears in 23 rate as well,
such that 
\begin{equation}
\frac{R_{23(HTL)}}{R_{23(MD)}}\simeq \delta .
\end{equation}%
On the other hand, the LPM effect is estimated to suppress the 23 rate by a
factor of 
\begin{equation}
\gamma =\frac{R_{23(LPM)}}{R_{23(MD)}}.
\end{equation}

Combining the estimated HTL and LPM corrections to the 23 rate, the 22+23
rate is likely to be in the range $[R_{22}+R_{23},R_{22}+\gamma \delta
R_{23}]$, while the higher order effect gives $\pm \sqrt{\alpha _{s}}\left(
R_{22}+R_{23}\right) $ corrections to the rate. Without further information,
the errors are assumed to be Gaussian and uncorrelated, the total rate is 
\begin{equation}
\left( R_{22}+\frac{\gamma \delta +1}{2}R_{23}\right) \pm \left( \frac{%
\gamma \delta -1}{2}R_{23}\right) \pm \sqrt{\alpha _{s}}\left(
R_{22}+R_{23}\right) ,
\end{equation}%
and the recommended upper ($\zeta _{+}$) and lower ($\zeta _{-}$) range for $%
\zeta $ are 
\begin{equation}
\zeta _{\pm }=\frac{1}{\left( R_{22}+\frac{\gamma \delta +1}{2}R_{23}\right)
\mp \sqrt{\left( \frac{\gamma \delta -1}{2}R_{23}\right) ^{2}+\alpha
_{s}\left( R_{22}+R_{23}\right) ^{2}}}.
\end{equation}%
The $\zeta _{\pm }$ values are shown in the left panel of Fig. \ref%
{fig:bulk23}.\

\end{document}